\newif\ifrelease
\releasetrue %

\newif\ifdiff
\difffalse %

\newif\ifextended
\extendedtrue

\newif\ifFINDINGSBOXES
\FINDINGSBOXESfalse
\FINDINGSBOXEStrue

\PassOptionsToPackage{dvipsnames}{xcolor}
\ifrelease
    \ifdiff
        \documentclass[sigconf, screen, review]{acmart}
    \else
        \ifextended
            \documentclass[sigconf, screen, nonacm]{acmart}
        \else
            \documentclass[sigconf, screen]{acmart}
        \fi
    \fi
\else
    \documentclass[sigconf, screen, review]{acmart}
\fi

\PassOptionsToPackage{frozencache}{minted}

\usepackage[T1]{fontenc}		%
\usepackage[linesnumbered,ruled,vlined]{algorithm2e}   %
\usepackage[nolist,nohyperlinks]{acronym}
\usepackage[skins,minted]{tcolorbox}  %
\usepackage[utf8]{inputenc}		%
\usepackage{array}  %
\usepackage{enumitem}       %
\usepackage{fontawesome5}   %
\usepackage{graphicx}
\usepackage{hyperref}
\usepackage[capitalise]{cleveref}
\crefname{figure}{Figure}{Figures} %
\crefalias{section}{appendix}
\usepackage{listings}
\usepackage{makecell}       %
\usepackage{mdframed} %
\usepackage{multirow}
\usepackage{subcaption}
\usepackage[dvipsnames]{xcolor}
\usepackage{tikz}  %
\usepackage{xspace}
\usepackage{xurl}		%
\usepackage{natbib}
\usepackage{longtable}
\usepackage{environ}

\graphicspath{{Figures/}}

\ifrelease
    \usepackage[disable]{todonotes} %
\else
    \usepackage[]{todonotes} %
\fi

\usepackage{soul}           %

\usepackage{fmtcount} %

\usetikzlibrary{
    positioning,
    patterns,
    calc,
    fit}

\fvset{fontsize=\footnotesize,xleftmargin=8pt,numbers=left,numbersep=5pt,frame=single}

\begin{acronym}
\acro{PTM}{pre-trained model}
\acro{AST}{abstract syntax tree}
\acro{CPG}{code property graph}
\acro{PM}{Pickle Machine}

\end{acronym}

\newcommand{\para}[1]{\noindent\hspace{0.15cm}\textbf{\textit{#1:}}}

\newcommand{\ie}{i.e.,\xspace}

\newcommand{\eg}{e.g.,\xspace}

\newcommand{\etal}{\textit{et al.}\xspace}

\newcommand{\scode}[1]{{\small \texttt{#1}}}

\newcommand{\ignore}[1]{}

\def\addauthnote#1#2{%
	\expandafter\def\csname#1\endcsname##1{%
		\todo[size=\footnotesize,color=#2,inline]
			{\textbf{\underline{\texttt{#1}}:} ##1}\xspace}
   \expandafter\def\csname#1li\endcsname##1{%
		\todo[size=\footnotesize,color=#2,inline,inlinewidth=5cm, noinlinepar]
			{\textbf{\underline{\texttt{#1}}:} ##1}\xspace}
    \expandafter\def\csname#1Res\endcsname##1{%
		\todo[size=\footnotesize,color=gray,inline]
			{\textbf{\underline{[RESOLVED] \texttt{#1}}:} ##1}\xspace}
}

\newcommand{\mintedfsize}{\footnotesize}
\newcommand{\mintedfont}{courier}
\definecolor{mintedbackground}{rgb}{0.95, 0.95, 0.95}
\definecolor{mintedframe}{rgb}{0.70,0.85,0.95}
\setminted[python3]{
	fontsize=\mintedfsize,
	fontfamily=\mintedfont,
	bgcolor=mintedbackground,
	escapeinside=||,
	linenos,
	breaklines,
	xleftmargin=4pt,
	tabsize=2
}
\newtcblisting{myminted}[2][]{
	enhanced,
	listing engine=minted,
	listing only,#1,
	title=#2,
	minted language=python3,
	minted options={python3},
	fonttitle=\ttfamily\footnotesize,
	colback=mintedbackground,
}

\newmintinline[inlPython]{python3}{fontsize=\small, escapeinside=||}
\newmintinline[inlPythonfn]{python3}{fontsize=\footnotesize, escapeinside=||}

\newcommand{\tool}{{\textsc{Pickle\-Ball}}\xspace}
\def\ptitle{\tool: Secure Deserialization of Pickle-based Machine Learning Models}
\def\bptitle{\tool: Secure Deserialization of Pickle-based \\ Machine Learning Models (Extended Report)}
\def\pkeywords{Secure Model Loading; Deserialization Attacks; Supply Chains}

\newcommand{\modelscan}{\mbox{\textsc{ModelScan}}\xspace}
\newcommand{\modeltracer}{\mbox{\textsc{ModelTracer}}\xspace}
\hypersetup{
	colorlinks	= true,
	citecolor	= {red!70!black},
	filecolor	= black,
	linkcolor	= {green!70!black},
	urlcolor	= {blue!70!black},
	pagebackref	= false,
	pdftitle	= {\ptitle},
	pdfkeywords	= {\pkeywords}
}

\addauthnote{vpk}{blue!25}
\addauthnote{neo}{green!25}
\addauthnote{andreas}{orange!25}
\addauthnote{yaniv}{red!25}
\addauthnote{jamie}{purple!25}
\addauthnote{wenxin}{green!50}
\addauthnote{laurent}{yellow!25}
\addauthnote{junfeng}{cyan!25}
\addauthnote{penghui}{red!25}
\addauthnote{TODO}{yellow!50}

\makeatletter
\if@todonotes@disabled

\else

\fi
\makeatother

\ifFINDINGSBOXES
    \newcommand{\summarybox}[1]{%
    \begin{tcolorbox}[width=\linewidth, colback=yellow!10!white, top=1pt, bottom=1pt, left=2pt, right=2pt]
    #1
    \end{tcolorbox}}
\else
    \newcommand{\summarybox}[1]{}
\fi

\setlist[itemize]{leftmargin=*}
\setlist[enumerate]{leftmargin=*}

\newcommand{\pickle}{pickle\xspace}

\newcommand{\Pickle}{Pickle\xspace}

\newcommand{\PM}{\ac{PM}\xspace}
\newcommand{\picklemachine}{Pickle Machine\xspace}
\newcommand{\weightsonly}{weights-only unpickler\xspace}

\newcommand{\NJS}{Node.js\xspace}

\newcommand{\allowedimports}{\emph{allowed imports}\xspace}
\newcommand{\allowedinvocations}{\emph{allowed invocations}\xspace}
\newcommand{\Allowedimports}{\textbf{Allowed imports}\xspace}
\newcommand{\Allowedinvocations}{\textbf{Allowed invocations}\xspace}
\newcommand{\queue}{\emph{candidate queue}\xspace}
\newcommand{\mockobjects}{stub objects\xspace}
\newcommand{\mockobject}{stub object\xspace}

\newcommand{\AST}{\ac{AST}\xspace}

\NewEnviron{NoDiff}{\BODY}

\newcommand{\pickleprevalencestudy}{44.9}
\newcommand{\pickleprevalencecasey}{41}
\newcommand{\pickleprevalencemalhug}{55}

\newcommand{\pickleprevalencestudypickleonlyrepos}{4,553\xspace}

\xspace
\newcommand{\tracedRepositories}{1,426\xspace}
\newcommand{\tracedRepositoriesDisallowed}{219\xspace}
\newcommand{\tracedRepositoriesDisallowedRate}{15.4\%\xspace} %
\newcommand{\disallowedrepositoriesdownloads}{{79.6 million}\xspace}
\newcommand{\tracedUnusualCallables}{36\xspace}

\newcommand{\ntotaldatasetmodels}{336\xspace} %
\newcommand{\nmaliciousdatasetmodels}{84\xspace} %
\newcommand{\nmaliciousdatasetmodelsreal}{82\xspace}
\newcommand{\nmaliciousdatasetmodelssynthetic}{2\xspace}
\newcommand{\nsafedatasetmodels}{\modelsTotal\xspace}
\newcommand{\ndatasetlibraries}{16\xspace}

\newcommand{\conchStars}{342}
\newcommand{\conchDownloads}{13.7K}
\newcommand{\flagembeddingStars}{9.3K}
\newcommand{\flagembeddingDownloads}{11.1M}
\newcommand{\flairStars}{14.1K}
\newcommand{\flairDownloads}{3.05M}
\newcommand{\glinerStars}{1.9K}
\newcommand{\glinerDownloads}{760K}
\newcommand{\huggingsoundStars}{447}
\newcommand{\huggingsoundDownloads}{56.1M}
\newcommand{\languagebindStars}{800}
\newcommand{\languagebindDownloads}{495K}
\newcommand{\melottsStars}{5.9k}
\newcommand{\melottsDownloads}{406K}
\newcommand{\parrotStars}{890}
\newcommand{\parrotDownloads}{911K}
\newcommand{\pyannoteStars}{7.2k}
\newcommand{\pyannoteDownloads}{32.6M}
\newcommand{\pysentimientoStars}{588}
\newcommand{\pysentimientoDownloads}{1.31M}
\newcommand{\sentencetransformersStars}{16.4k}
\newcommand{\sentencetransformersDownloads}{204M}
\newcommand{\superimageStars}{170}
\newcommand{\superimageDownloads}{64.9K}
\newcommand{\tnerStars}{387}
\newcommand{\tnerDownloads}{25.0K}
\newcommand{\tweetnlpStars}{341}
\newcommand{\tweetnlpDownloads}{80.7K}
\newcommand{\yolovfiveStars}{53.4k}
\newcommand{\yolovfiveDownloads}{24.8K}
\newcommand{\yolovelevenStars}{39.2k}
\newcommand{\yolovelevenDownloads}{38.4M}

\newcommand{\conchImportsObserved}{3}
\newcommand{\conchImportsAllowed}{822}
\newcommand{\conchImportsStubbed}{0}
\newcommand{\conchInvocationsObserved}{2}
\newcommand{\conchInvocationsAllowed}{61}
\newcommand{\conchInvocationsStubbed}{0}
\newcommand{\conchModelsTotal}{1}
\newcommand{\conchModelsLoaded}{1}
\newcommand{\conchModelsSuccessrate}{100.0\%}

\newcommand{\flagembeddingImportsObserved}{4}
\newcommand{\flagembeddingImportsAllowed}{773}
\newcommand{\flagembeddingImportsStubbed}{0}
\newcommand{\flagembeddingInvocationsObserved}{2}
\newcommand{\flagembeddingInvocationsAllowed}{61}
\newcommand{\flagembeddingInvocationsStubbed}{0}
\newcommand{\flagembeddingModelsTotal}{14}
\newcommand{\flagembeddingModelsLoaded}{14}
\newcommand{\flagembeddingModelsSuccessrate}{100.0\%}

\newcommand{\flairImportsObserved}{34}
\newcommand{\flairImportsAllowed}{1186}
\newcommand{\flairImportsStubbed}{17}
\newcommand{\flairInvocationsObserved}{6}
\newcommand{\flairInvocationsAllowed}{62}
\newcommand{\flairInvocationsStubbed}{2}
\newcommand{\flairModelsTotal}{18}
\newcommand{\flairModelsLoaded}{17}
\newcommand{\flairModelsSuccessrate}{94.4\%}

\newcommand{\glinerImportsObserved}{3}
\newcommand{\glinerImportsAllowed}{870}
\newcommand{\glinerImportsStubbed}{0}
\newcommand{\glinerInvocationsObserved}{2}
\newcommand{\glinerInvocationsAllowed}{61}
\newcommand{\glinerInvocationsStubbed}{0}
\newcommand{\glinerModelsTotal}{17}
\newcommand{\glinerModelsLoaded}{17}
\newcommand{\glinerModelsSuccessrate}{100.0\%}

\newcommand{\huggingsoundImportsObserved}{3}
\newcommand{\huggingsoundImportsAllowed}{767}
\newcommand{\huggingsoundImportsStubbed}{0}
\newcommand{\huggingsoundInvocationsObserved}{2}
\newcommand{\huggingsoundInvocationsAllowed}{61}
\newcommand{\huggingsoundInvocationsStubbed}{0}
\newcommand{\huggingsoundModelsTotal}{17}
\newcommand{\huggingsoundModelsLoaded}{17}
\newcommand{\huggingsoundModelsSuccessrate}{100.0\%}

\newcommand{\languagebindImportsObserved}{4}
\newcommand{\languagebindImportsAllowed}{992}
\newcommand{\languagebindImportsStubbed}{0}
\newcommand{\languagebindInvocationsObserved}{2}
\newcommand{\languagebindInvocationsAllowed}{61}
\newcommand{\languagebindInvocationsStubbed}{0}
\newcommand{\languagebindModelsTotal}{8}
\newcommand{\languagebindModelsLoaded}{8}
\newcommand{\languagebindModelsSuccessrate}{100.0\%}

\newcommand{\melottsImportsObserved}{3}
\newcommand{\melottsImportsAllowed}{852}
\newcommand{\melottsImportsStubbed}{0}
\newcommand{\melottsInvocationsObserved}{2}
\newcommand{\melottsInvocationsAllowed}{61}
\newcommand{\melottsInvocationsStubbed}{0}
\newcommand{\melottsModelsTotal}{8}
\newcommand{\melottsModelsLoaded}{6}
\newcommand{\melottsModelsSuccessrate}{75.0\%}

\newcommand{\pyannoteImportsObserved}{18}
\newcommand{\pyannoteImportsAllowed}{1085}
\newcommand{\pyannoteImportsStubbed}{9}
\newcommand{\pyannoteInvocationsObserved}{5}
\newcommand{\pyannoteInvocationsAllowed}{64}
\newcommand{\pyannoteInvocationsStubbed}{0}
\newcommand{\pyannoteModelsTotal}{14}
\newcommand{\pyannoteModelsLoaded}{10}
\newcommand{\pyannoteModelsSuccessrate}{71.4\%}

\newcommand{\pysentimientoImportsObserved}{4}
\newcommand{\pysentimientoImportsAllowed}{777}
\newcommand{\pysentimientoImportsStubbed}{0}
\newcommand{\pysentimientoInvocationsObserved}{2}
\newcommand{\pysentimientoInvocationsAllowed}{61}
\newcommand{\pysentimientoInvocationsStubbed}{0}
\newcommand{\pysentimientoModelsTotal}{4}
\newcommand{\pysentimientoModelsLoaded}{4}
\newcommand{\pysentimientoModelsSuccessrate}{100.0\%}

\newcommand{\sentencetransformersImportsObserved}{5}
\newcommand{\sentencetransformersImportsAllowed}{1087}
\newcommand{\sentencetransformersImportsStubbed}{0}
\newcommand{\sentencetransformersInvocationsObserved}{2}
\newcommand{\sentencetransformersInvocationsAllowed}{61}
\newcommand{\sentencetransformersInvocationsStubbed}{0}
\newcommand{\sentencetransformersModelsTotal}{76}
\newcommand{\sentencetransformersModelsLoaded}{76}
\newcommand{\sentencetransformersModelsSuccessrate}{100.0\%}

\newcommand{\superimageImportsObserved}{3}
\newcommand{\superimageImportsAllowed}{1016}
\newcommand{\superimageImportsStubbed}{0}
\newcommand{\superimageInvocationsObserved}{2}
\newcommand{\superimageInvocationsAllowed}{61}
\newcommand{\superimageInvocationsStubbed}{0}
\newcommand{\superimageModelsTotal}{6}
\newcommand{\superimageModelsLoaded}{6}
\newcommand{\superimageModelsSuccessrate}{100.0\%}

\newcommand{\tweetnlpImportsObserved}{4}
\newcommand{\tweetnlpImportsAllowed}{778}
\newcommand{\tweetnlpImportsStubbed}{0}
\newcommand{\tweetnlpInvocationsObserved}{2}
\newcommand{\tweetnlpInvocationsAllowed}{61}
\newcommand{\tweetnlpInvocationsStubbed}{0}
\newcommand{\tweetnlpModelsTotal}{1}
\newcommand{\tweetnlpModelsLoaded}{1}
\newcommand{\tweetnlpModelsSuccessrate}{100.0\%}

\newcommand{\yolovfiveImportsObserved}{28}
\newcommand{\yolovfiveImportsAllowed}{920}
\newcommand{\yolovfiveImportsStubbed}{7}
\newcommand{\yolovfiveInvocationsObserved}{4}
\newcommand{\yolovfiveInvocationsAllowed}{61}
\newcommand{\yolovfiveInvocationsStubbed}{0}
\newcommand{\yolovfiveModelsTotal}{12}
\newcommand{\yolovfiveModelsLoaded}{4}
\newcommand{\yolovfiveModelsSuccessrate}{33.3\%}

\newcommand{\yolovelevenImportsObserved}{63}
\newcommand{\yolovelevenImportsAllowed}{1816}
\newcommand{\yolovelevenImportsStubbed}{13}
\newcommand{\yolovelevenInvocationsObserved}{11}
\newcommand{\yolovelevenInvocationsAllowed}{61}
\newcommand{\yolovelevenInvocationsStubbed}{6}
\newcommand{\yolovelevenModelsTotal}{51}
\newcommand{\yolovelevenModelsLoaded}{15}
\newcommand{\yolovelevenModelsSuccessrate}{29.4\%}

\newcommand{\parrotImportsObserved}{3}
\newcommand{\parrotImportsAllowed}{774}
\newcommand{\parrotImportsStubbed}{0}
\newcommand{\parrotInvocationsObserved}{2}
\newcommand{\parrotInvocationsAllowed}{61}
\newcommand{\parrotInvocationsStubbed}{0}
\newcommand{\parrotModelsTotal}{1}
\newcommand{\parrotModelsLoaded}{1}
\newcommand{\parrotModelsSuccessrate}{100.0\%}

\newcommand{\tnerImportsObserved}{4}
\newcommand{\tnerImportsAllowed}{769}
\newcommand{\tnerImportsStubbed}{0}
\newcommand{\tnerInvocationsObserved}{2}
\newcommand{\tnerInvocationsAllowed}{61}
\newcommand{\tnerInvocationsStubbed}{0}
\newcommand{\tnerModelsTotal}{4}
\newcommand{\tnerModelsLoaded}{4}
\newcommand{\tnerModelsSuccessrate}{100.0\%}

\newcommand{\modelsTotal}{252}
\newcommand{\modelsTotalPickleballLoaded}{201}
\newcommand{\modelsTotalSuccessRate}{79.8\%}
\newcommand{\modelsAvgPickleball}{87.7\%}
\newcommand{\modelsTotalPickleballFailed}{51}
\newcommand{\modelsTotalFailureRate}{20.2\%}

\newcommand{\conchModelsWOU}{1}
\newcommand{\conchWOUSuccessrate}{100.0\%}
\newcommand{\flagembeddingModelsWOU}{14}
\newcommand{\flagembeddingWOUSuccessrate}{100.0\%}
\newcommand{\flairModelsWOU}{0}
\newcommand{\flairWOUSuccessrate}{0.0\%}
\newcommand{\glinerModelsWOU}{17}
\newcommand{\glinerWOUSuccessrate}{100.0\%}
\newcommand{\huggingsoundModelsWOU}{17}
\newcommand{\huggingsoundWOUSuccessrate}{100.0\%}
\newcommand{\languagebindModelsWOU}{8}
\newcommand{\languagebindWOUSuccessrate}{100.0\%}
\newcommand{\melottsModelsWOU}{8}
\newcommand{\melottsWOUSuccessrate}{100.0\%}
\newcommand{\pyannoteModelsWOU}{0}
\newcommand{\pyannoteWOUSuccessrate}{0.0\%}
\newcommand{\pysentimientoModelsWOU}{4}
\newcommand{\pysentimientoWOUSuccessrate}{100.0\%}
\newcommand{\sentencetransformersModelsWOU}{76}
\newcommand{\sentencetransformersWOUSuccessrate}{100.0\%}
\newcommand{\superimageModelsWOU}{6}
\newcommand{\superimageWOUSuccessrate}{100.0\%}
\newcommand{\tweetnlpModelsWOU}{1}
\newcommand{\tweetnlpWOUSuccessrate}{100.0\%}
\newcommand{\yolovfiveModelsWOU}{0}
\newcommand{\yolovfiveWOUSuccessrate}{0.0\%}
\newcommand{\yolovelevenModelsWOU}{0}
\newcommand{\yolovelevenWOUSuccessrate}{0.0\%}
\newcommand{\parrotModelsWOU}{1}
\newcommand{\parrotWOUSuccessrate}{100.0\%}
\newcommand{\tnerModelsWOU}{4}
\newcommand{\tnerWOUSuccessrate}{100.0\%}

\newcommand{\modelsTotalWOULoaded}{157}
\newcommand{\modelsTotalWOUSuccessrate}{62.3\%}
\newcommand{\modelsAvgWOU}{75.0\%}
\newcommand{\modelsTotalWOUFailed}{95}
\newcommand{\modelsTotalWOUFalureRate}{37.7\%}

\newcommand{\timeGenerationMinValueSec}{9.0\xspace}
\newcommand{\timeGenerationMinLibrary}{CONCH\xspace}
\newcommand{\timeGenerationMaxValueSec}{29.8\xspace}
\newcommand{\timeGenerationMaxLibrary}{YOLOv11\xspace}
\newcommand{\timeGenerationMedian}{14.0\xspace}

\newcommand{\runtimeEnforcementOverheadMs}{0.42ms\xspace}
\newcommand{\runtimeEnforcementOverheadPer}{1.75\%\xspace}
\newcommand{\runtimeEnforcementAvgOverheadPer}{2.62\%\xspace}

\newcommand{\libraryconch}{CONCH\xspace}
\newcommand{\libraryflagembedding}{FlagEmbedding\xspace}
\newcommand{\libraryflair}{flair\xspace}
\newcommand{\librarygliner}{GLiNER\xspace}
\newcommand{\libraryhuggingsound}{huggingsound\xspace}
\newcommand{\librarylanguagebind}{LanguageBind\xspace}
\newcommand{\librarymelotts}{MeloTTS\xspace}
\newcommand{\libraryparrot}{Parrot\_Paraphraser\xspace}
\newcommand{\librarypyannote}{PyAnnote\xspace}
\newcommand{\librarypysentimiento}{pysentimiento\xspace}
\newcommand{\librarysentencetransformers}{sentence\_transformers\xspace}
\newcommand{\librarysuperimage}{super-image\xspace}
\newcommand{\librarytner}{TNER\xspace}
\newcommand{\librarytweetnlp}{tweetnlp\xspace}
\newcommand{\libraryyolovfive}{YOLOv5\xspace}
\newcommand{\libraryyoloveleven}{YOLOv11 (ultralytics)\xspace}

\newcommand{\libraryconchhash}{\texttt{02d6ac5}}
\newcommand{\libraryflagembeddinghash}{\texttt{bf6b649}}
\newcommand{\libraryflairhash}{\texttt{c674212}}
\newcommand{\libraryglinerhash}{\texttt{1169120}}
\newcommand{\libraryhuggingsoundhash}{\texttt{50e9fba}}
\newcommand{\librarylanguagebindhash}{\texttt{7070c53}}
\newcommand{\librarymelottshash}{\texttt{5b53848}}
\newcommand{\libraryparrothash}{\texttt{03084c5}}
\newcommand{\librarypyannotehash}{\texttt{0ea4c02}}
\newcommand{\librarypysentimientohash}{\texttt{60822ac}}
\newcommand{\librarysentencetransformershash}{\texttt{a1db32d}}
\newcommand{\librarysuperimagehash}{\texttt{50439ea}}
\newcommand{\librarytnerhash}{\texttt{7730a62}}
\newcommand{\librarytweetnlphash}{\texttt{68b08c8}}
\newcommand{\libraryyolovfivehash}{\texttt{40a1887}}
\newcommand{\libraryyolovelevenhash}{\texttt{b18007e}}

\newcommand{\libraryconchrepo}{mahmoodlab/CONCH}
\newcommand{\libraryflagembeddingrepo}{FlagOpen/FlagEmbedding}
\newcommand{\libraryflairrepo}{flairNLP/flair}
\newcommand{\libraryglinerrepo}{urchade/GLiNER}
\newcommand{\libraryhuggingsoundrepo}{jonatasgrosman/huggingsound}
\newcommand{\librarylanguagebindrepo}{PKU-YuanGroup/LanguageBind}
\newcommand{\librarymelottsrepo}{myshell-ai/MeloTTS}
\newcommand{\libraryparrotrepo}{PrithivirajDamodaran/Parrot\_Paraphraser}
\newcommand{\librarypyannoterepo}{pyannote/pyannote-audio}
\newcommand{\librarypysentimientorepo}{pysentimiento/pysentimiento}
\newcommand{\librarysentencetransformersrepo}{UKPLab/sentence-transformers}
\newcommand{\librarysuperimagerepo}{eugenesiow/super-image}
\newcommand{\librarytnerrepo}{asahi417/tner}
\newcommand{\librarytweetnlprepo}{cardiffnlp/tweetnlp}
\newcommand{\libraryyolovfiverepo}{fcakyon/yolov5-pip}
\newcommand{\libraryyolovelevenrepo}{ultralytics/ultralytics}

\input{Figures/fmt} %

\begin{document}

\copyrightyear{2025}
\acmYear{2025}
\setcopyright{acmlicensed}\acmConference[CCS '25]{Proceedings of the 2025 ACM SIGSAC Conference on Computer and Communications Security}{October 13--17, 2025}{Taipei, Taiwan}
\acmBooktitle{Proceedings of the 2025 ACM SIGSAC Conference on Computer and Communications Security (CCS '25), October 13--17, 2025, Taipei, Taiwan}
\acmDOI{10.1145/3719027.3765037}
\acmISBN{979-8-4007-1525-9/2025/10}

\title[\ptitle]{\bptitle}

\author{Andreas D. Kellas}
\orcid{0009-0001-8763-2400}
\affiliation{
    \institution{Columbia University}
    \city{New York}
    \state{NY}
    \country{USA}
}
\email{andreas.kellas@columbia.edu}

\author{Neophytos Christou}
\orcid{0000-0001-7335-9485}
\affiliation{
    \institution{Brown University}
    \city{Providence}
    \state{RI}
    \country{USA}
}
\email{neophytos_christou@brown.edu}

\author{Wenxin Jiang}
\orcid{0000-0003-2608-8576}
\affiliation{
    \institution{Purdue University}
    \city{West Lafayette}
    \state{IN}
    \country{USA}
}
\additionalaffiliation{
    \institution{Socket}
    \city{Wilmington}
    \state{DE}
    \country{USA}
}
\email{jiang784@purdue.edu}

\author{Penghui Li}
\orcid{0000-0002-3077-5697}
\affiliation{
    \institution{Columbia University}
    \city{New York}
    \state{NY}
    \country{USA}
}
\email{pl2689@columbia.edu}

\author{Laurent Simon}
\orcid{0000-0001-7893-547X}
\affiliation{
    \institution{Google}
    \city{Mountain View}
    \state{CA}
    \country{USA}
}
\email{laurentsimon@google.com}

\author{Yaniv David}
\orcid{0000-0002-1630-7723}
\affiliation{
    \institution{Technion}
    \city{Haifa}
    \country{Israel}
}
\email{yanivmd@technion.ac.il}

\author{Vasileios P. Kemerlis}
\orcid{0000-0002-6528-437X}
\affiliation{
    \institution{Brown University}
    \city{Providence}
    \state{RI}
    \country{USA}
}
\email{vpk@cs.brown.edu}

\author{James C. Davis}
\orcid{0000-0003-2495-686X}
\affiliation{
    \institution{Purdue University}
    \city{West Lafayette}
    \state{IN}
    \country{USA}
}
\email{davisjam@purdue.edu}

\author{Junfeng Yang}
\orcid{0009-0000-2277-6545}
\affiliation{
    \institution{Columbia University}
    \city{New York}
    \state{NY}
    \country{USA}
}
\email{junfeng@cs.columbia.edu}

\renewcommand{\shortauthors}{Andreas D. Kellas et al.}

\begin{abstract}

Machine learning model repositories, such as the Hugging Face Model Hub, facilitate model exchanges.
However, bad actors can deliver malware through compromised models.
Existing defenses, such as safer model formats, restrictive (but inflexible) loading policies, and model scanners, have shortcomings:
  \pickleprevalencestudy\% of popular models on Hugging Face still use the
  insecure \pickle format,
  15\% of these cannot be loaded by restrictive loading policies,
  and model scanners have both false positives and false negatives.
\Pickle remains the \textit{de facto} standard for model exchange, and the ML
community lacks a tool that offers transparent safe loading.

We present \tool to help machine learning engineers load \pickle-based models safely.
\tool statically analyzes the source code of machine learning libraries
and computes custom policies that specify a safe load-time behavior for benign
models.
It then dynamically enforces
these policies during load time as a drop-in replacement for the \pickle module.
\tool generates policies that correctly load \modelsTotalSuccessRate~of benign
\pickle-based models in our dataset, while rejecting all (100\%) malicious
examples in the same dataset.
In comparison, evaluated model scanners fail to identify known malicious models,
and the state-of-the-art loader loads 22\% fewer benign models than \tool.
\tool removes the threat of arbitrary function invocation from
malicious \pickle-based models, raising the bar for attackers as they have to depend
on code reuse techniques.

\end{abstract}


\maketitle

\section{Introduction}\label{SEC:INTRO}

Open-source and open-weight models enable the AI
ecosystem~\cite{jiang2022ptmsupplychain,davis2023reusing,wang2022PTLMandApplications,han2021PTMs}.
They allow machine learning engineers to exchange pre-trained models rather
than training from scratch~\cite{dubey2024llama3}, facilitating direct use
or fine-tuning~\cite{church2021emergingTrendsFinetuning}.
Open model repositories like Hugging Face now host millions of pre-trained
models for many
tasks~\cite{jiang2023ptmreuse, jiang2022ptmsupplychain}.
These model hubs are accessed directly as well as through corporate
mirrors~\cite{zhao2024MalHug}, with billions of downloads per month.

Similar to traditional software supply-chain attacks, bad actors can
distribute malicious models.
The most common strategy for achieving remote code execution is tampering
with the model deserialization process.
Several model serialization formats, such as
TorchScript~\cite{torchscript}, H5/HDF5~\cite{hdf5}, and the Python \pickle
module~\cite{pickle-docs}, permit executable metadata or callbacks during model
deserialization.
Attackers can craft malicious serialized models to execute code, such as
\texttt{os.system()}, on victim systems during model
loading~\cite{casey2024largescaleexploitinstrumentationstudy,jfrog_malicious_models_detected,hiddenlayer_machine_learning_threat_roundoup,
sleepy_pickle_exploit}.
Researchers have found malicious \pickle models on Hugging
Face whose payloads include system fingerprinting, credential theft, and reverse shells~\cite{%
  jfrog_malicious_models_detected,%
  hiddenlayer_machine_learning_threat_roundoup,%
  zhao2024MalHug%
},
with one study discovering a $5{\times}$ increase in the rate of malicious
models uploaded to Hugging Face year-over-year~\cite{zhao2024MalHug}.

This led to alternative safe model formats like
SafeTensors~\cite{huggingface_safetensors}, and restricted loading APIs like the
PyTorch \weightsonly~\cite{weightsOnlyUnpickler}; we study their adoption in the
Hugging Face ecosystem (\S\ref{SEC:MOTIVATION}) and find that nevertheless,
insecure formats are still prevalent.

In this work, we propose a novel approach to secure \pickle model
deserialization, which we focus on for three reasons.
First, \pickle is a popular exchange format for models.
Repositories with \pickle models are downloaded over 2.1 billion times per month
from the Hugging Face model hub (\S\ref{SEC:MOTIVATION_SURVEY}).
Second, \pickle is the most expressive format and thus is challenging to secure.
Models are encoded as opcodes that are executed by the \pickle virtual
machine~\cite{fickling, huang2022painpickle} during deserialization, which
permits invocations of arbitrary Python classes and functions
(\emph{callables}).
Third, \pickle is abused by attackers.
Almost all malicious models on Hugging Face use the \pickle format.

Our evaluation shows that the two existing kinds of \pickle deserialization
defenses are inadequate.
\textit{Model scanners}~\cite{%
  protectai_modelscan,%
  huggingface_picklescanning,%
  zhao2024MalHug,%
  casey2024largescaleexploitinstrumentationstudy%
}
maintain fixed denylists of disallowed callables to identify models that invoke
them.
\textit{Safe model loaders}~\cite{weightsOnlyUnpickler} use fixed allowlists to
permit only the use of trusted callables. %
Our evaluation shows the limits of these inflexible approaches for ML models.
For instance, the default safe deserialization loader in
PyTorch~\cite{weightsOnlyUnpickler} prevents 15\% of Hugging Face \pickle
repositories from loading (\S\ref{SEC:MOTIVATION_SURVEY}).

To address these limitations, we present \tool, a two-part system for securing
the exchange of \pickle-based models.
Our insight is that we can analyze library code to determine the expected
behaviors of benign models produced by the library, and enforce
\textit{tailored model loading policies}.
\tool statically analyzes the library code to learn
  (1)~all class types used in the library's models and their transitive attribute types,
  and
  (2)~all functions needed to restore objects of these types.
Then, \tool's model-loading component enforces the inferred policies.

We evaluated \tool and the state-of-the-art approaches on a dataset of
\ntotaldatasetmodels models.
Our dataset is meant to represent the kinds of models that \tool must handle,
including malicious and benign models.
We used \nsafedatasetmodels benign models sourced from Hugging Face, and
\nmaliciousdatasetmodels real and synthetic malicious models.
\tool loads \modelsTotalSuccessRate~of benign models
and prevents all malicious models from executing their
payloads.
\tool adds a runtime overhead of $\sim$\runtimeEnforcementAvgOverheadPer to
safely load a model.
Compared to other approaches, \tool achieved favorable precision and recall. %

In summary, we contribute:

\begin{enumerate}

    \item An \textbf{ecosystem-scale study} of \pickle security considerations
    in the
    Hugging Face Model
    Hub.
    Repositories with \textit{only} \pickle models are downloaded over 400
    million times per month, despite the introduction of new model formats.
    We find that 15\% of repositories with only \pickle models have a model
    that cannot be loaded by the SOTA secure \pickle model loader, the \weightsonly.

    \item \tolerance=1600
    The \textbf{design and implementation of \tool}, a
    framework for securely loading \pickle models.
    \tool tailors loading policies to models, and enforces these policies
    lazily, for secure, efficient, and robust model loading.%
    \footnote{\tool is available at \url{https://github.com/columbia/pickleball}.}

    \item A \textbf{novel dataset} of \ntotaldatasetmodels benign and malicious
    \pickle-based models for evaluating \pickle security efforts.
    It has
      \nsafedatasetmodels benign models collected from Hugging Face,
      and
      \nmaliciousdatasetmodels malicious models.\!%

\end{enumerate}

\section{Background} \label{SEC:BACKGROUND}

Here we
  describe the ML model reuse ecosystem and formats
  (\S\ref{SEC:BACKGROUND_MODELUSECASES}),
 then
  how the \pickle format is used and the risks it entails
  (\S\ref{SEC:BACKGROUND_PICKLE}).

\subsection{Model Reuse} \label{SEC:BACKGROUND_MODELUSECASES}

\subsubsection{The Model Supply Chain}

Training models from scratch requires significant resources%
~\cite{patterson2021carbon, dubey2024llama3},
so engineers and companies reuse machine-learning models trained by others
(pre-trained models)%
~\cite{jiang2022ptmsupplychain,davis2023reusing,wang2022PTLMandApplications,han2021PTMs}.
Open model repositories like Hugging Face host over 1.8M
models~\cite{huggingface-data} for many tasks~\cite{jiang2023ptmreuse,
jiang2022ptmsupplychain}.

A supply chain of pre-trained models has grown from model reuse,
which comes with risks similar to those in the traditional software
supply chain~\cite{ohm2020backstabber, jiang2022ptmsupplychain}.
Bad actors apply techniques familiar in traditional software security, such as
typosquatting~\cite{neupane2024beyondtyposquatting,liu2022containertyposquatting,jiang2024naming,jiang2025ConfuGuard}
and code injection~\cite{wang2022evilmodel}, as well as ML-specific techniques
like model and data manipulation~\cite{gu2019badnets}.
Model hubs like Hugging Face try to detect malicious models using both traditional
code scanners like ClamAV~\cite{jiang2022ptmsupplychain} and
domain-specific \pickle
scanners~\cite{protectai_modelscan,huggingface_picklescanning} due to the
proliferation of malicious \pickle
models~\cite{casey2024largescaleexploitinstrumentationstudy,zhao2024MalHug}.

Machine learning libraries facilitate the development and exchange of models.
Libraries like PyTorch~\cite{pytorch},
TensorFlow~\cite{tensorflow2015-whitepaper}, and JAX~\cite{jax2018github}
provide a core of general ML library utilities, like model training and
serialization functions.
Other popular but more specific libraries build upon the core libraries with
task-specific utilities, like ultralytics (formerly YOLO)~\cite{yolov11} for
image recognition, PyAnnote~\cite{pyannote} for audio processing, and
flair~\cite{flair} for text processing.
To create a model, an engineer uses one of these libraries to write a \textit{model
saver} program, which trains a model and serializes it for reuse.
To load and use the model, an engineer uses the same library (identified in
documentation that accompanies the model) to write a \textit{model loader}
program.
The libraries provide the interface for interacting with the shared models.

\subsubsection{Model Serialization Formats}

Model savers and loaders must agree on the serialization format; there are
various formats available, each with its own
tradeoffs in terms of security, flexibility, and performance.%
  \!~\footnote{
    A table showing these tradeoffs is
    provided in the SafeTensors repository \texttt{README}:
    \url{https://github.com/huggingface/safetensors}.}
Python is the primary language for using ML models, and its native serialization
module, \pickle~\cite{pickle-docs}, provides a convenient and flexible
interface for saving objects;
  \pickle proliferated for being easy to use and is used by popular
  libraries like PyTorch~\cite{pytorch-serialization}.
Hugging Face released the SafeTensors format in September 2022 as an alternative
that prioritizes security~\cite{huggingface_safetensors}.
The GGUF format, released August 2023, is optimized for fast model loading and
inference tasks, especially for large language
models~\cite{gguf_github_readme,gguf_huggingface}.
Other formats may be selected for library coupling (\eg~TensorFlow SavedModel),
interoperability (\eg ONNX), or intermediate tradeoffs between security,
flexibility, and performance.

The security of a format depends on the expressivity of its operations.
The SafeTensors format requires very few different operations to load a model,
because it only encodes model weight values, and is considered a
secure format after independent security
audits~\cite{trailofbitsSafeTensorAudit,huggingface_safetensor_security_audit}.
Some formats, like TensorFlow SavedModel and ONNX, are known to permit
operations that could be abused in specific settings%
~\cite{zhu:2025:mymodelismalwaretoyou,onnx_runtime_hacks},
but with no observed real-world attacks.

\Pickle is an extremely expressive format that permits nearly arbitrary
operations during deserialization, and numerous malicious \pickle models are
discovered on Hugging Face%
~\cite{zhao2024MalHug,%
casey2024largescaleexploitinstrumentationstudy,%
reversinglabs_malicious_models_discovered,%
jfrog_malicious_models_detected,%
hiddenlayer_machine_learning_threat_roundoup}.
We focus our efforts on \pickle models due to their insecure format (cf.~\S\ref{SEC:BACKGROUND_PICKLE}) and their continued popularity (cf.~\S\ref{SEC:MOTIVATION}).

\subsection{Pickle Serialization and Risks}\label{SEC:BACKGROUND_PICKLE}

\Pickle is popular because of its flexibility and convenience, due to its
ability to represent almost any Python object.
The \pickle module implements a virtual machine, the \textit{\ac{PM}}, that
executes a sequence of opcodes~\cite{thePVM} to deserialize an object.
The expressiveness of the \ac{PM} allows it to serialize and reconstruct complex
Python data structures, but also make it vulnerable to attacks, allowing
attackers to invoke arbitrary Python callables~\cite{fickling}.

\para{Pickle Program Structure and Semantics}
A \pickle program consists of opcode sequences interpreted by the \ac{PM}, a
stack-based VM implemented in the Python \pickle module~\cite{fickling}.
When the \pickle program halts, the object at the top of the \ac{PM} stack is
returned to the Python interpreter as the deserialized object.

\ignore{%
that is interpreted during the model loading process. Typically, the \pickle
program is used to read model weight values from files and instantiate the
model architecture, and so is packaged with other model data, often in a zip
archive, depicted in Figure~\ref{FIG:PICKLE_BASED_MODEL_STRUCTURE}.

\Pickle programs are interpreted to initialize the serialized object that they
encode. They contain sequences of opcodes that are interpreted by the \ac{PM}, a
stack-based virtual machine embedded in the Python interpreter. When the \pickle
program halts, the object at the top of the \ac{PM} stack is returned to the
Python interpreter as the deserialized object.}

\begin{figure}[t]
    \input{Figures/motivation.py.tex}
  \caption{Example of a machine learning library with a model that can
  be pickled. The \inlPython{Tensor} class's
  \inlPython{__reduce__} method registers the \inlPython{read_weights_to_tensor}
  function for execution during deserialization.}
  \label{LISTING:LIBRARYPY}
\end{figure}

The \ac{PM} is integrated into the Python interpreter. %
Many of the \ac{PM}'s opcodes create or manipulate native-type objects,
\eg~\texttt{NEW\-FALSE} to create a \texttt{bool}, while others import and invoke
Python \emph{callables} (classes and functions)~\cite{python-callables}.
Specifically, the \emph{callable importing} opcodes \texttt{GLOBAL} and
\texttt{STACK\_GLOBAL} take a callable's name, import it, and push it to the top
of the \ac{PM} stack.
Class instances are instantiated using \emph{callable allocating} opcodes, like
\texttt{NEWOBJ}, which calls the class's \inlPython{__new__} method.
Function references are called via \emph{callable invoking} opcodes like
\texttt{REDUCE}.
Arguments can be passed to both allocations and invocations, and the return
value is pushed to the \PM stack.
Lastly, the \emph{callable building}
\texttt{BUILD} opcode can modify an object (\eg~set/change its
attributes).
\Pickle lets users customize the deserialization process
with the \inlPython{__reduce__} method.

The method must return a reference to a function and arguments.
During serialization, an object's \inlPython{__reduce__} method is called and
the returned function reference and argument values are written to the
serialized output opcodes.
During deserialization, the function is invoked with the arguments, and
the return value is pushed onto the \ac{PM} stack.
This provides a primitive to invoke arbitrary functions in a \pickle program.
\cref{LISTING:LIBRARYPY} shows an example class that registers a handler
function (line 10) that is invoked with a \emph{callable invoking} opcode during
deserialization.

\para{Pickle Deserialization Attacks} \Pickle's opcodes
allow a \pickle program to \emph{import} and \emph{invoke}
arbitrary Python callables during unpickling. An attacker can
use of \pickle's \emph{callable importing} and \emph{callable invoking}
opcodes to achieve arbitrary code execution---for example by executing an
arbitrary shell command by invoking the \texttt{os.system} function. The
dangers of deserializing arbitrary \pickle programs have been
publicized since 2011~\cite{defcontalk, fickling, nelhage,pickle-docs}.

\para{Protecting Pickle Models}
Two existing approaches are used to protect users from malicious \pickle models:
\begin{enumerate}
  \item \textbf{Model scanners} identify malicious models using denylists of
  unsafe callables.
  As with many denylist approaches, model scanners are useful for identifying
  recognizable malicious content, but are bypassed by malicious models that
  avoid denied callables, or that \emph{indirectly} invoke
  callables~\cite{checkmarx:bypassing_scanners}.\!%
    \footnote{We demonstrate this by creating
    two backdoored models that bypass two state-of-the-art scanners. One model
    uses callables that are missed by the scanners, and the other model uses
    disallowed callables by invoking them indirectly. See
    \cref{SEC:MOTIVATION_SCANNERS}.}
  Examples are Hugging Face's picklescan~\cite{huggingface_picklescanning}
  and ProtectAI's scanner~\cite{huggingface_protectai_announcement}.

  \item \textbf{Restricted loaders} restrict the \ac{PM} to execute only allowed,
  safe callables.
  The only available restricted \pickle model loader is PyTorch's
  \weightsonly~\cite{weightsOnlyUnpickler}.
  Its default allowlist is
  specialized for models produced by PyTorch.
\end{enumerate}

\section{Motivation}\label{SEC:MOTIVATION}

To summarize Section~\ref{SEC:BACKGROUND}: it is dangerous to
load untrusted \pickle models, but alternative secure formats exist.
Do \pickle models remain a security threat?
We answer this with
  a longitudinal study of \pickle model usage (\S\ref{SEC:MOTIVATION_SURVEY}),
  and assessments of the usability of the PyTorch \weightsonly
  (\S\ref{SEC:MOTIVATION_WEIGHTSONLY}),

\subsection{Study of Pickle Models on Hugging Face}\label{SEC:MOTIVATION_SURVEY}

To determine whether \pickle models are used despite the availability of
alternate formats, we conducted a longitudinal study of the Hugging Face
ecosystem.
We investigated Hugging Face because it is
  the largest repository of pre-trained models~\cite{jiang2022ptmsupplychain},
  and
  because it hosts malicious \pickle models~\cite{%
    casey2024largescaleexploitinstrumentationstudy,%
    zhao2024MalHug,%
    jfrog_malicious_models_detected,%
    hiddenlayer_machine_learning_threat_roundoup,%
    reversinglabs_malicious_models_discovered}.
At 10 points in time over a $\sim$2-year period, we measured the download rates
and model formats in repositories with $\ge\!1000$ monthly downloads (as a proxy for
real-world impact).\footnote{Downloads are tracked using
\href{https://huggingface.co/docs/hub/en/models-download-stats}{Hugging Face
metrics}.}
The number of repositories in a measurement ranged from 2,296 in the first
measurement to 16,661 in the last, with the number increasing monotonically at
each point.
We mined two existing datasets that covered January--October 2023  (PTMTorrent~\cite{jiang2023ptmtorrent} and HFCommunity~\cite{ait2023hfcommunity}),
  and added new measurements in August 2024, November 2024, and March 2025 via the \texttt{huggingface\_hub} API.
In accordance with previous research~\cite{zhao2024MalHug}, we determined model
formats using file extensions; interested readers can refer
to \cref{HFMeasureAPdix} for details.

\begin{figure}
    \centering
    \includegraphics[width=0.99\linewidth]{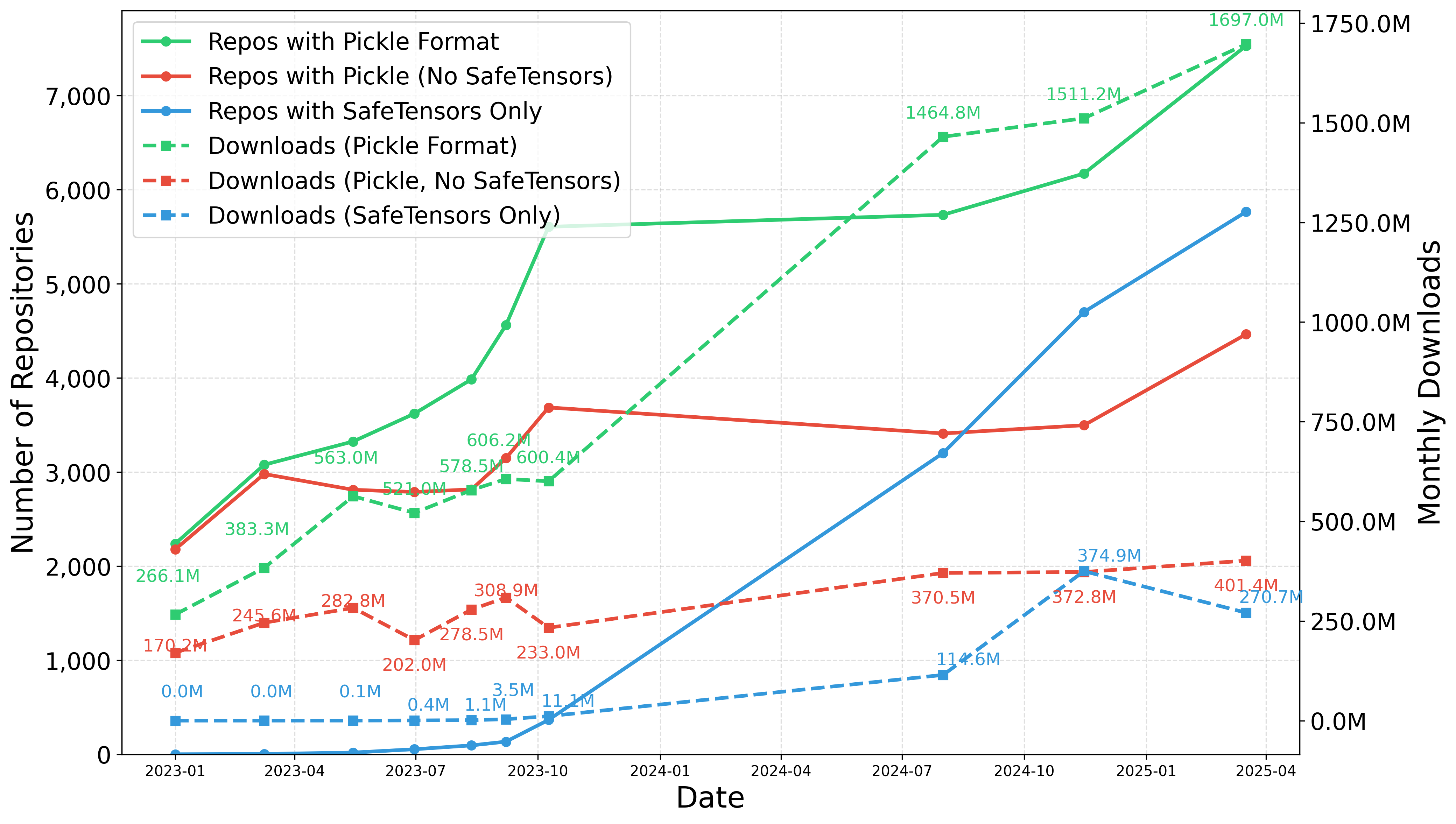}
    \caption{
    A longitudinal analysis of Hugging Face model formats for repositories with
    $\ge\!1000$ monthly downloads.
    Repositories can contain multiple models, each in different formats.
    Each color groups repositories by the model formats they contain: at least
    one \pickle model (green), exclusively \pickle (red), and exclusively
    SafeTensors (blue).
    }
    \label{fig:survey}
\end{figure}

\begin{figure}
  \centering
  \includegraphics[width=0.99\linewidth]{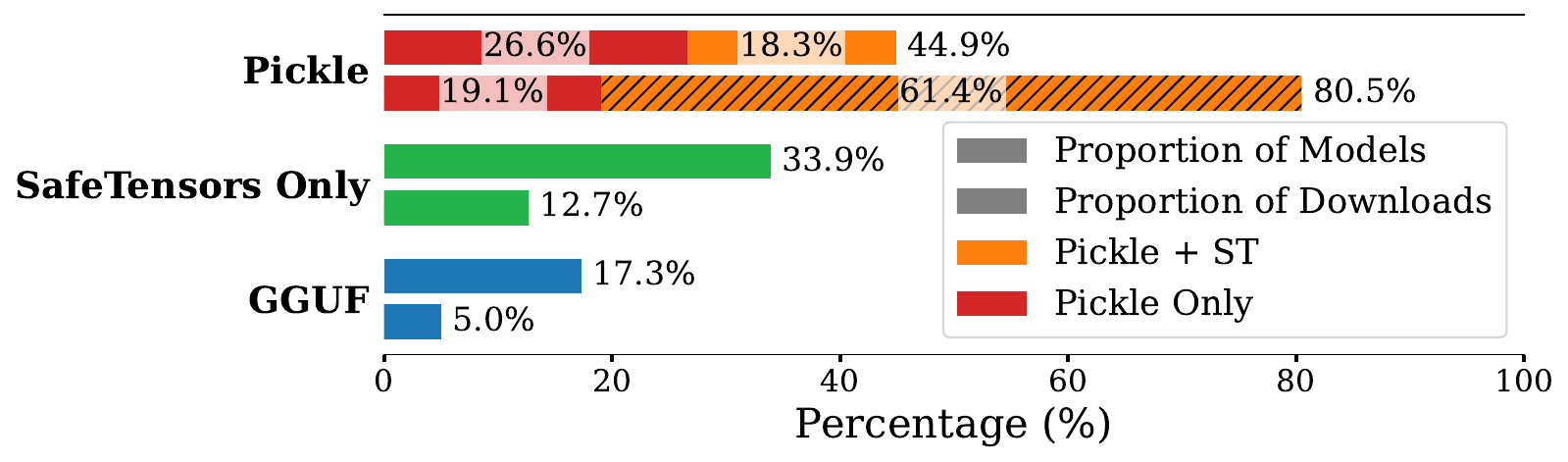}
  \caption{
  Proportions of model formats and downloads in March 2025, the final month of
  our longitudinal study.
  Notation: ``Pickle+ST'' indicates repositories with both formats.
  }
  \label{fig:surveyfinalmonth}
\end{figure}

\cref{fig:survey} summarizes our results.
First, the red lines show that many important models continue to use only the
\pickle format, and these \pickle-only models are downloaded 400M+ times per
month.
The green lines show that repositories containing both \pickle and SafeTensors
versions of models are also increasingly downloaded, with 1.70 billion monthly
downloads.
When models are converted to the SafeTensors format, the associated \pickle
model is often kept for legacy purposes and can still present security
risks~\cite{llama_31_pickle}.

\cref{fig:surveyfinalmonth} represents present-day usage with data from
the final month of our study.
Repositories with only SafeTensors or GGUF models are downloaded infrequently,
in comparison to those with \pickle models.
Overall, $\sim$\pickleprevalencestudy\% of repositories contain \pickle models,
  which aligns with previous estimates of
  \pickleprevalencecasey\%--\pickleprevalencemalhug\%%
  ~\cite{%
    casey2024largescaleexploitinstrumentationstudy,%
    zhao2024MalHug}.

We anticipate that \pickle models will continue to pose risks to the Hugging
Face community for the next few years (cf.~\S\ref{SEC:DISCUSSION}).
Monthly download rates of \pickle models are \emph{increasing}, and
many (21\%) models are still exclusively in the \pickle format, including
  29 models in the top-100 most downloaded
  and
  over 500 models from Meta, Google, Microsoft, NVIDIA, and Intel.
PyTorch remains the primary framework for model development, which reinforces
reliance on \pickle due to user familiarity~\cite{tan2022DLSupplyChain}.
Interoperability challenges persist during model
conversion~\cite{trailofbits2023security, jajal2024interoperability, hf-blog1},
complicating movement to other formats.

\summarybox{\textbf{Summary:}
Despite positive steps to introduce secure alternative model formats like
SafeTensors, \pickle models are still prevalent and monthly downloads are
\emph{increasing}.}

\subsection{PyTorch Weights-Only Unpickler Usability}\label{SEC:MOTIVATION_WEIGHTSONLY}

Our longitudinal study showed that \pickle remains popular. Next, we assess
whether the state-of-the-art safe loading approach, the PyTorch \weightsonly,
can effectively load the \pickle models we identify.

\subsubsection{Measurements}
The \weightsonly, introduced in PyTorch 1.13 (Nov. 2022) and enabled by
default in PyTorch 2.6 (Nov. 2024)~\cite{weights-only-default}, prevents
access to dangerous callables, but can only load models that use callables from a
small allowed set of PyTorch APIs.\footnote{PyTorch provides a mechanism for the user to manually expand the set
  of allowed callables~\cite{weights-only-addglobals}, but the user is left to
  determine by themselves
  \textit{which} callables to allow.}
Models that use additional callables cannot be loaded without user intervention;
convenience and pressure from end-users results in library maintainers
explicitly disabling the \weightsonly to maintain compatibility.%
\footnote{As in the case of the \texttt{flairNLP} library (\url{https://github.com/flairNLP/flair/commit/79aa33706e7f753f2edf962feb1d75de22af0d1d}).}
We investigate a sample of the \pickle models in our study to determine whether
they use callables disallowed by the \weightsonly, which affects the usability
of the \weightsonly as a solution.

\para{Methods}
We sampled the most popular 1,500 of the \pickleprevalencestudypickleonlyrepos
\pickle-only repositories in our survey (\S\ref{SEC:MOTIVATION_SURVEY}).
For each repository, we used the Hugging Face API to download its \pickle
models and used the fickling tool~\cite{fickling} to statically trace and
inspect the callables used.
We compared the callables in the model trace to the callables permitted by the
\weightsonly's default policy.
Models from 74 repositories failed to download or trace, leaving us with a
sample of \tracedRepositories repositories.

\para{Results}
Of the \tracedRepositories model repositories surveyed,
\emph{\tracedRepositoriesDisallowed repositories
(\tracedRepositoriesDisallowedRate) contained at least one \pickle-based model
that cannot be loaded by the \weightsonly} due to disallowed Python callables.
These \tracedRepositoriesDisallowed repositories were downloaded
\disallowedrepositoriesdownloads~times in the final month of our longitudinal
study.
In total, \tracedUnusualCallables unique disallowed callables appear in the
\tracedRepositoriesDisallowed repositories.
Many come from major libraries (\eg numpy and Hugging Face Transformers).
We list all disallowed callables that appear in traces in
\cref{FIG:UNUSUAL_CALLABLES} (\cref{HFMeasureAPdix}).
These models cannot be loaded securely by the \weightsonly, so users must
instead rely on the weaker model scanners (\S\ref{SEC:BACKGROUND_PICKLE}) and
their own assessments of the models' safety.

\subsubsection{Motivating Example}\label{SEC:MOTIVATION_EXAMPLE}

\begin{figure}[t]
  \centering
  \includegraphics[scale=0.20]{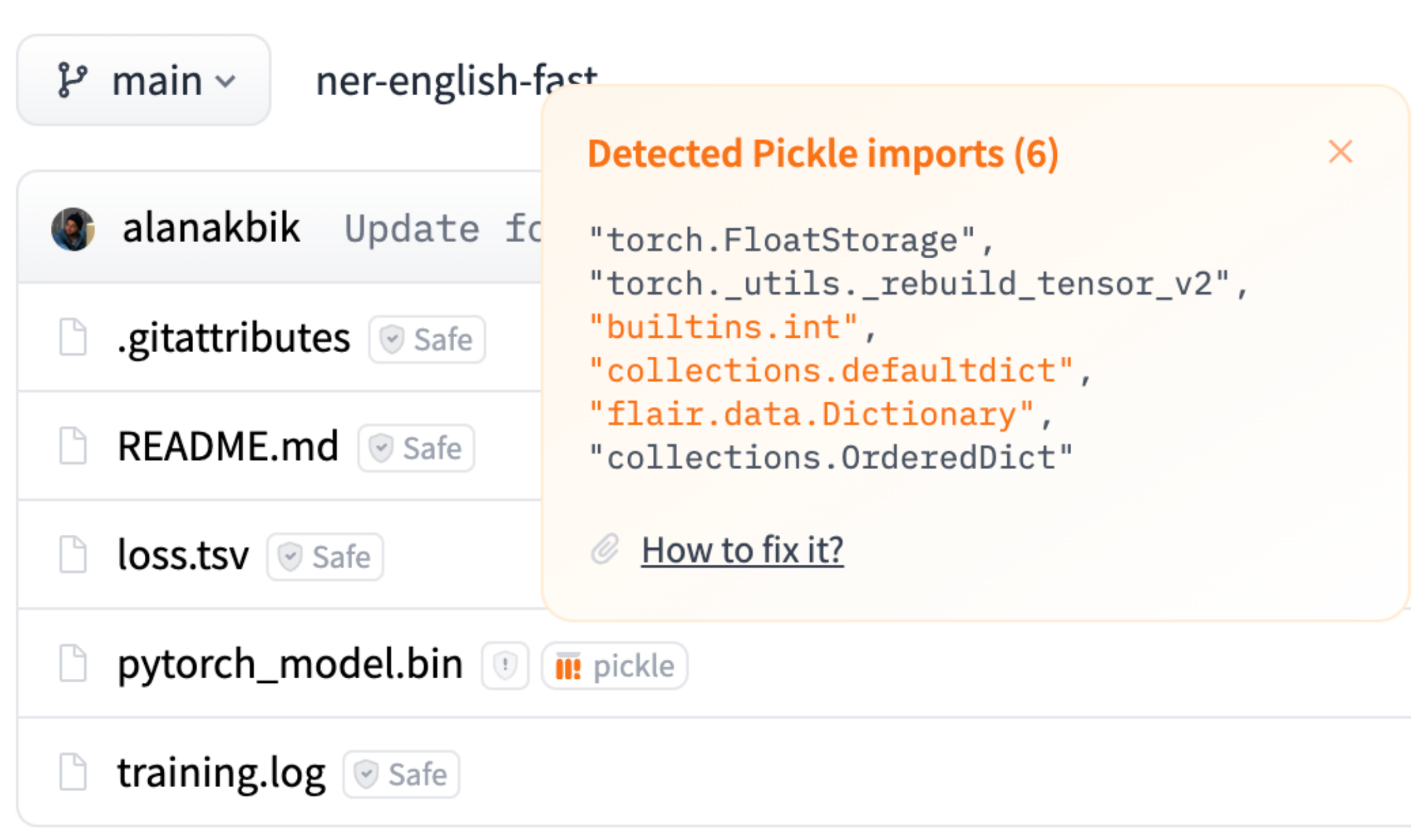}
    \caption{
  The Hugging Face repository for the \texttt{flair\textbackslash ner-english-fast} model
  shows the results of the Hugging Face \pickle scanning tool directly in the web
  application interface~\cite{flair_ner_english_fast_model}. The \pickle scanning
  tool warns that some imports in the model's \pickle file are suspicious and
  require attention (highlighted).
  }
    \label{FIG:MOTIVATING_EXAMPLE_PYANNOTE}
\end{figure}

We use an example to show the implications of these \weightsonly
incompatibilities. Consider the
\texttt{flair} \texttt{ner-english-fast}~\cite{flair_ner_english_fast_model} model, a pre-trained pickle model for named entity recognition of
English text that gained over 1 million downloads. %
To load the model, its documentation refers to the  %
\texttt{flair} library's \inlPython{SequenceTagger.load}\xspace API.

The \texttt{flair} loading API exposes the user to risk.
\texttt{Flair} models, like this one, use callables that are not part of the
\weightsonly allowlist, so the API explicitly disables the \weightsonly to
load its models.
To protect themselves during loading, the user must depend on scanners.
For this (benign) model, the Hugging Face \pickle
scanner~\cite{huggingface_picklescanning} warns the user that some imports in
the model \pickle file are suspicious and require attention
(\cref{FIG:MOTIVATING_EXAMPLE_PYANNOTE}).
We determined that the model is benign by manually reviewing its operations, and then reviewing the source code of the flair library to ensure that
these operations are expected.
This task is costly for every user to perform for every model.
Our system successfully infers the expected operations and securely loads this
model (\S\ref{SEC:EVAL_SAFE_LOADING}) without manual effort.

\summarybox{
\textbf{Summary}:
While the \weightsonly offers security, \tracedRepositoriesDisallowedRate of
sampled \pickle repositories, with \disallowedrepositoriesdownloads monthly
downloads, contain a model that cannot use it. Library APIs disable the
\weightsonly to load models, leaving users to rely on (incomplete) model
scanners and manual assessments to determine if models are malicious.}

\section{System and Threat Model}\label{SEC:MOTIVATION_THREAT_MODEL}

Our motivational study shows the need for a new defense that is both usable and secure.
Here, we model the system we
aim to protect and the adversary to thwart:

\para{System Model} The system loads a \pickle-based model
from an untrusted source (\eg~Hugging Face Model Hub) using APIs provided by a
trusted ML library (\eg~PyTorch).
We specifically aim to protect the system from the code introduced by the \pickle program and executed by the \picklemachine.

\para{Threat Model} The attacker provides a
maliciously crafted \pickle to the victim with the intention of
compromising the system.
The attacker's goal is to execute arbitrary Python code (a ``\textit{payload}''), either directly during
model loading, or after by \eg overwriting a method in the model
object with a reference to the payload.

\begin{itemize}

\item \textit{In scope:} Manipulation of a \pickle program
in a \pickle-based serialized model to execute arbitrary code.

\item \tolerance=1600 \textit{Out of scope:} Manipulation of the data or
code in the serialized model beyond the \pickle program (\eg~model weights, data pipeline programs); manipulation of the trusted library code (\eg~PyTorch) used to load the serialized model.

\end{itemize}

We focus on the threat of \pickle program code execution, and exclude other
threats from untrusted ML models.
Other ML supply chain attacks, like manipulating model weights to insert
``backdoors''~\cite{gu2019badnets,impnet}, are orthogonal and can be approached
with layered defenses.
We do not consider other forms of attacks that manipulate \pickle programs,
\eg for denial of service~\cite{fickling_dos_issue} due to their weaker attack
primitives.

\section{\tool Design and Implementation} \label{SEC:OVERVIEW}

\begin{figure}[t]
    \centering
    \resizebox{\columnwidth}{!}{
    \input{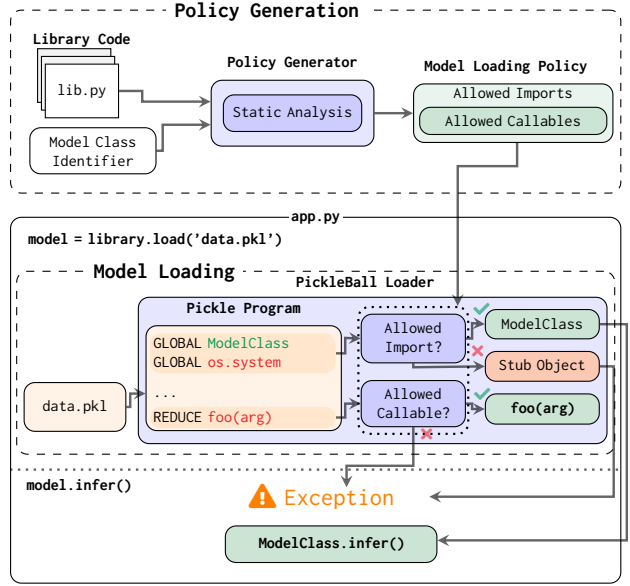}
    }
    \caption{\tool works in two phases: 1)~policy generation and 2)~safe model
    loading. During policy generation, \tool takes as input the source code of a
    ML library and a class definition to analyze, and outputs a policy of
    allowed imports and invocations. During safe model loading, \tool
    enforces the extracted policy to protect the loading process. The
    loading application specifies the policy to enforce, based on the expected
    class of the model, and begins loading the model with the library API. The
    loading application can trust that any invocations of the \picklemachine
    will be restricted to the configured policy.
    }
    \label{FIG:OVERVIEW}
\end{figure}

\tool is designed to protect applications that use libraries to load untrusted
\pickle models.
The desired system guarantee is that \tool raises a security exception when an
adversary invokes an unnecessary callable during model loading, while
transparently loading benign models.
The idea behind \tool is to first generate a policy describing a minimal set of
operations (\ie~the callables that need to be imported and invoked) for
instantiating a given model object
(\S\ref{SEC:OVERVIEW_STATIC}), and then to enforce the generated policy during
unpickling, rejecting spurious operations performed by malicious models
(\S\ref{SEC:OVERVIEW_DYNAMIC}).
\cref{FIG:OVERVIEW} provides an overview of the aforementioned \tool components.
\tool guarantees that, given correct AST and type information, \tool raises a
security exception when the adversary invokes an unnecessary callable
(\S\ref{SEC:OVERVIEW_GUARANTEES}). We implement the design of \tool as a
software artifact (\S\ref{SEC:IMPLEMENTATION}).

\subsection{Policy Generation}\label{SEC:OVERVIEW_STATIC}

\tool's policy generation component is designed to automatically create a policy
that describes the set of operations (\ie the imported and invoked callables)
permitted when loading a model of a particular library class.
The policy is generated before loading the untrusted \pickle model, and restricts
the loading behaviors to only those that are necessary for the library API.

\para{Design Rationale}
We guide our design by studying how \pickle models make use of Python callables.
In regular usage, a \pickle program needs a callable to construct non-primitive objects
and (recursively) initialize the values of its attributes.
Callables that are not needed to instantiate a given object should
not appear in a \pickle program; malicious payloads insert new code that does not
have a role in object initialization.

The challenging task of policy generation is determining \textit{which}
callables are needed to instantiate an object.
Python is a dynamic language that permits a single variable to receive
different types at various program paths, and for objects to receive new
attribute variables after initialization.
Identifying all attribute types requires a path-sensitive analysis of the
object creation code up to the point that the object is serialized.
Further, in the ML setting, the code that creates the model object is not
provided: the model saving program that trained the model is not published along
with the model.

However, we recognize that we can generate approximate policies for serialized
models by analyzing the class definition of the model in the library source
code, even without access to the model saving program.
The majority of the class instance attribute information is contained in the
class definition, rather than in the model saving application.
Intuitively, for the object to be reusable, it has to conform to an expected
interface that is defined in the library.

We hypothesize that any unobserved object writes after the model is created
will either not introduce new types to the variables, or will describe
specialized metadata that is not necessary for the general-use operations that
the model loading application is likely to perform, like inference.
If new types are introduced, the model class interface shared between the saving
and loading programs is violated.
This is supported by our evaluation (\S\ref{SEC:EVAL_BENIGN_RESULTS}).

\para{Policy Generation Algorithm}
Given an ML library and model class definition, \tool analyzes the class
definition to generate a model loading policy as the sets of
\allowedimports and \allowedinvocations; these represent the operations
that a \pickle program needs to instantiate an instance of the class:

\begin{enumerate}
    \item \Allowedimports: the set of callables permitted as arguments to
    \picklemachine import operations.
    \item \Allowedinvocations: the set of callables permitted
    arguments to \picklemachine invoking operations.
\end{enumerate}

The set of \allowedinvocations is a sub-set of \allowedimports---before being
invoked, the callable must be imported. However, a callable that is imported but
not invoked may only be used as a reference or as a constructor with its
allocator method (\inlPython{__new__}).

To statically generate the policy for a given class, \tool implements and
applies rules starting at the class definition, and proceeds until the analysis
terminates.
\tool maintains a \queue of classes that is initialized with the class
definition.
\tool adds new classes to the queue as they are discovered by the analysis
rules, and removes them as each class is analyzed.
A class is only added to the candidate queue once to ensure that the analysis
terminates.

\tool applies the following class-analysis rules:
\begin{enumerate}
\item If the class implements a \inlPython{__reduce__} method: identify the method return values
    (a callable, arguments for the callable, and optional state initialization
    values); add the returned callable to the \allowedimports and \allowedinvocations
    sets; identify the types of all arguments for the callable and state
    initialization values, and add their class definitions to the \queue.
\item Otherwise: add the class to the \allowedimports set; add all
    sub-classes of the class to the \queue; add all types of the class's
    attributes (including attributes inherited) to the \queue.
\end{enumerate}

The \tool policy generation algorithm, shown in \cref{ALGO:STATIC_INFERENCE},
operates over an \AST representation of the analyzed program, and
expects recovered type information to be labeled in the \AST.

\tool's static analysis cannot produce perfectly sound and precise policies
due to fundamental challenges in statically analyzing Python code, which is
dynamically typed~\cite{xu:2016:pythonprobabilistictypeinference,aycock:2000:aggressivetypeinference,gorbovitski:2010:aliasanalysis}.
When recovered type information is over-approximate, \tool produces policies
that contain more callables than is strictly necessary.
Under-approximate type information may result from Python's dynamic features,
like dynamic typing and runtime attribute manipulation, producing unaccounted
data dependencies during type recovery.
The missing type information creates policies that incorrectly exclude
callables.
To account for these potential errors, we design for security by separately
enforcing allowed import and allowed invocations, and for robustness with a
lazy enforcement mechanism
(\S\ref{SEC:OVERVIEW_DYNAMIC}).
We discuss how static analysis limitations affect the whole-system analysis in
\cref{SEC:OVERVIEW_GUARANTEES}.

\newcommand{\AlgGets}{$\;:=\;$}
\newcommand{\AlgPGets}{$\mathrel{{+}{=}}$\xspace}

\newcommand{\Candidates}{Candidates}
\newcommand{\Candidate}{Candidate}
\newcommand{\AllowedGlobals}{AllowedImports}
\newcommand{\AllowedReduces}{AllowedInvocations}
\newcommand{\Superclass}{SuperClass}

\SetKwFunction{Set}{Set}
\SetKwFunction{Queue}{UniqueQueue}
\SetKwFunction{EmptySet}{EmptySet}
\SetKwFunction{NotEmpty}{NotEmpty}
\SetKwFunction{Pop}{Pop}
\SetKwFunction{HasReduceMethods}{HasReduceMethods}
\SetKwFunction{GetReduceMethodReturn}{GetReduceReturn}
\SetKwFunction{GetReduceMethodArgumentTypes}{GetReduceReturnTypes}
\SetKwFunction{GetSubclasses}{GetSubclasses}
\SetKwFunction{GetAttributeTypes}{GetAttributeTypes}

\begin{algorithm}
\captionsetup{width=\textwidth}
\caption{
The pseudocode algorithm designed to generate a model loading policy for a
given class, performed over an \AST augmented with recovered type information.
Errors in the recovered type information introduce incorrectness in the
results of \texttt{GetReduceReturnTypes} and \texttt{GetAttributeTypes}.
}
\label{ALGO:STATIC_INFERENCE}
\small
\KwIn{ModelClass}

\KwOut{\AllowedGlobals}
\KwOut{\AllowedReduces}

\Candidates \AlgGets \Queue{ModelClass} \\
\AllowedGlobals \AlgGets \EmptySet \\
\AllowedReduces \AlgGets \EmptySet \\

\While{\NotEmpty{\Candidates}}{
    \Candidate \AlgGets \Pop{\Candidates} \\
    \If{\HasReduceMethods{\Candidate}}{
        \AllowedGlobals \AlgPGets \GetReduceMethodReturn{\Candidate} \\
        \AllowedReduces \AlgPGets \GetReduceMethodReturn{\Candidate} \\
        \Candidates \AlgPGets \GetReduceMethodArgumentTypes{\Candidate} \\
    }\Else{
        \AllowedGlobals \AlgPGets \Candidate \\
        \Candidates \AlgPGets \GetSubclasses{\Candidate} \\
        \Candidates \AlgPGets \GetAttributeTypes{\Candidate} \\
    }
}

\end{algorithm}

\subsection{Policy Enforcement}\label{SEC:OVERVIEW_DYNAMIC}

\tool's policy enforcement module is designed to protect the model loading
application during loading.
It is a drop-in replacement for the system \pickle module.
It restricts the behavior of the \ac{PM} operations that import,
initialize, and invoke Python callables, so that only the callables
permitted by the model loading policy are accessible to the \pickle program in
the model.

The module receives the loading policy and \pickle
program as inputs, and either outputs the deserialized object from the \pickle
program, or raises a security exception.
``Importing opcodes'' can only access callables that are allowed by the
\allowedimports policy.
``Allocating opcodes'' are allowed to create instances of objects listed in the
\allowedimports set by invoking their \inlPython{__new__} methods.
Callable-invoking opcodes are either removed entirely, or restricted to
only invoke callables that are in the \allowedinvocations.

To enforce this, the module verifies that the name of a given callable
is present in the allowed invocations set.
Conceptually, an attacker could bypass the allowed
imports/allowed invocations separation by importing a callable present in the
allowed imports set but not in the allowed invocations set, then ``renaming''
it to a callable that is in the allowed invocations set.
To mitigate against this, the policy enforcement module prevents building
opcodes from modifying
\inlPython{__name__} and \inlPython{__module__} attributes.

As described in Section~\ref{SEC:OVERVIEW_STATIC}, \tool policies may exclude
valid callables when \tool fails to analyze dynamic Python features;
\tool is designed to be robust against this with \textit{lazy enforcement}.
\tool aims to handle cases where the excluded callables are not accessed in the
model's downstream use cases.
When \tool's loader encounters a disallowed import operation, it creates a
\mockobject instead of immediately raising an exception.
The \mockobject records the name of the callable, and
implements no functionality other than raising a security exception when
invoked or accessed.
This permits the loader to proceed to completion in the event that the
\mockobject is never used, deferring the violation until access.
The security exception is also raised when the \mockobject is accessed after
initialization, preventing an attacker from overwriting methods of the returned
object with denied callables that are later invoked.

\subsection{Security Guarantees and Limitations}\label{SEC:OVERVIEW_GUARANTEES}

\para{Whole-system Guarantees}
\tool's design thwarts the adversary described in our Threat Model
(\S\ref{SEC:MOTIVATION_THREAT_MODEL}) by restricting the \ac{PM}'s access
to Python callables.
The design guarantees that when provided a correct \AST with type information
for a Python class definition, \tool (1) raises a security exception when the
adversary invokes a spurious callable, while (2) successfully instantiating any
object that respects the attribute and type information defined class
definition.
In our evaluation of \tool, we show that existing state-of-the-art
static analysis tools provide sufficient \AST and type-recovery
information for practical use (\S\ref{SEC:EVAL}).

\para{Policy Generation Guarantees}
When \tool is provided a correct \AST with type information, it is
guaranteed to output a loading policy that includes all \allowedimports and
\allowedinvocations that can appear in when an object is saved in the \pickle
format, provided that the object is not manipulated to add attribute types
outside of its object prototype.

\para{Policy Enforcement Guarantees}
When \tool loads a model, it is guaranteed to prevent the invocation
of any Python callable that is not in the configured set of \allowedinvocations,
and to create sanitized \mockobjects for any callable that is not in the
configured set of \allowedimports. The \mockobjects raise security exceptions
when accessed/\-invoked.

\para{Whole-system Limitations}
\tool is limited fundamentally by the challenges of analyzing dynamic Python
code with static analysis techniques, but \tool takes steps to mitigate these.
Python's dynamic features, like runtime attribute manipulation and dynamic
typing, prevent \tool's static analyses from creating an \AST with sound and
precise type information;
  this makes \tool's policies unsound and incomplete.
Over-approximations in the \AST result in policies that permit more Python
callables than necessary; \tool mitigates this by having separate
\allowedinvocations and \allowedimports policies, so that only a small
set of callables may be invoked.
Under-approximations in the \AST result in policies that omit benign callables
from valid models; \tool mitigates this with \textit{lazy policy enforcement} so
that omitted callables only raise exceptions if they are invoked by the model,
rather than just initialized but unused.
We evaluate \tool in Section~\ref{SEC:EVAL} to determine whether these limitations
restrict it in practical settings (and find that they do not).

\para{Remaining Attack Surface}
\tool prevents attackers from importing and
invoking arbitrary callables for malicious payloads.
However, akin to return-to-\texttt{libc} attacks, \tool does not prevent the
attacker from invoking permitted callables in sequences or with parameters that
result in unintended outcomes.
We are unaware of attacks leveraging these primitives, but whether this
remaining capability is exploitable is a subject for further
research~(\S\ref{SEC:DISCUSSION}).

\subsection{Implementation} \label{SEC:IMPLEMENTATION}

\tool is implemented in a total of ${\sim}$1,300 Scala lines of code (LoC) and
${\sim}$300 Python LoC divided between two primary components: a static program
analysis that builds upon the Joern framework~\cite{yamaguchi2014joern}; and a
dynamic loader that modifies the existing \picklemachine.
In the static analysis, ${\sim}$700 Scala LoC implement
\cref{ALGO:STATIC_INFERENCE}, ${\sim}$600 Scala LoC extend and fix Joern features,
and ${\sim}$200 Python LoC integrate components.
In the loader, ${\sim}$100 Python LoC modify the \picklemachine to implement
lazy policy enforcement.
Joern provides a program analysis platform for \tool by generating
a Code Property Graph (CPG) with recovered type information for the target
code; we extend Joern to improve type recovery features and class inheritance
tracking
(for more details, see \cref{ImpAPdix}).
Our analysis queries Joern's AST nodes for the relevant information.

\para{Limitations}
We inherit some limitations from the Joern program analysis framework.
The limitations include that \tool:
\begin{itemize}
\item Cannot parse new Python syntax features (\eg generic types).
\item Cannot recognize type hints provided in docstring comments, but it
    can process type annotations introduced in Python 3.5.
\item May fail to resolve dependencies, especially of builtin types.
\item Cannot identify attributes of classes implemented in C.
\end{itemize}
These are engineering limitations and can be addressed with improvements to the
underlying static analysis framework;
    they are not fundamental limitations of the \tool approach for determining
    model loading policies from library class definitions.

To account for these limitations when evaluating the \tool approach, we apply
some manual library pre-processing before analysis (\S\ref{SEC:EVAL_DATASET}),
and discuss future work to reduce the need for manual changes
(\S\ref{SEC:DISCUSSION}).

\section{Evaluation} \label{SEC:EVAL}

We evaluate \tool with four Research Questions (RQs):

\begin{itemize}

    \item \textbf{RQ1: Malicious Model Blocking.} How well does \tool block malicious
    pickled models from executing their payloads?

    \item \textbf{RQ2: Benign Model Loading.} How well does \tool correctly
    load benign pickled models?

    \item \textbf{RQ3: Performance.} Is \tool's runtime overhead practical for
    deployment?

    \item \textbf{RQ4: Comparison to SOTA.} How does \tool compare to the state-of-the-art
    security tools for protecting model loading applications?

\end{itemize}

\subsection{Constructing an Evaluation Dataset}\label{SEC:EVAL_DATASET}

To answer these questions, we need a comprehensive dataset of \pickle
models consisting of both malicious and benign examples.
The dataset must contain models created by different libraries to represent the
diversity of loading APIs.
We created our dataset by combining existing datasets, open models on
Hugging Face, and constructing synthetic models.
Our dataset contains \nsafedatasetmodels benign and \nmaliciousdatasetmodels malicious
models, for a total of \ntotaldatasetmodels.

\para{Benign Models and Trusted Libraries}
Our dataset must represent how users typically load and use models;
therefore, we need a set of benign models and the libraries used to
load and interact with them.
We first searched for libraries that meet three criteria:
(1) they load \pickle models;
(2) they have a model class type that is returned by a loading API; and
(3) if a foundational library (\eg~PyTorch or transformers) type is used, the
custom class adds new attributes to the type.
These criteria are motivated by \tool's purpose: to restrict the \pickle
operations permitted by a library loading API based on analysis of the intended
class type.

We identified candidate libraries by identifying popular \pickle models on Hugging
Face and working backwards.
We first searched Hugging Face programmatically for \pickle
models, ordered by monthly download rate, that had model loading documentation
directing users to a model loading library API.
We manually reviewed the top 400 models (approximately 2 hours of review time)
to determine whether the identified libraries meet our criteria; this resulted
in \ndatasetlibraries accepted libraries.
All libraries and their associated version information are listed in \cref{TABLE:library-versions},
\cref{SEC:APDX:DATASET}.

Then, we identified candidate models associated with each library.
We again queried Hugging Face to identify models associated with the library,
either directly (as a piece of repository metadata) or by mention in the model
documentation or name.
We collected models with $\ge$ 100 monthly downloads at time of collection.
In total, we accumulated \nsafedatasetmodels models produced by
\ndatasetlibraries different libraries.
All collected models are listed in \cref{TABLE:model-versions}, \cref{SEC:APDX:DATASET}.

We acknowledge that these models could themselves be malicious.
We partially mitigate this by sampling from the most frequently downloaded
models and libraries, checking model scanner indications, and manually
investigating unexpected callables when restricted model loaders identify them.

\para{Malicious Models}
Our dataset must also represent the models created by our intended adversary
(cf. our threat model --- \S\ref{SEC:MOTIVATION_THREAT_MODEL}); therefore, we
need a set of malicious \pickle models.
We first collected \nmaliciousdatasetmodelsreal malicious models and \pickle
programs that were identified on Hugging Face by two state-of-the-art model
scanners%
~\cite{casey2024largescaleexploitinstrumentationstudy, zhao2024MalHug}.
We add our \nmaliciousdatasetmodelssynthetic malicious models constructed to
bypass scanners (see \cref{SEC:MOTIVATION_SCANNERS}),
    for a total of \nmaliciousdatasetmodels.
All malicious models contain \pickle programs with payloads that %
import and invoke Python spurious functions;
    payload behaviors include accessing sensitive files, making network
    connections, creating reverse shells, among others.

We acknowledge that these models do not represent the complete set of malicious
behaviors;
    it is a best-effort collection of real-world \pickle model
    malware to represent today's attackers.
We aim to prevent adversaries that execute arbitrary Python functions during
\pickle loading, and the collected models all exercise this feature.

\para{Test Harnesses}\label{SEC:EVAL_DATASET_TEST_PROGRAMS}
\tool protects programs that load untrusted models; therefore, we need a
representative set of loading programs to secure.
We create one test harness for each library in our dataset; the harness loads a
model using the library API and performs an inference task.

\para{Library Pre-processing}
We pre-process the libraries before analyzing them to account for limitations
in the static analysis framework (\S\ref{SEC:IMPLEMENTATION}) and improve the
correctness of the AST.
We make manual source code modifications ($<$10 LoC) when
the library class uses newer Python features of Python that Joern's front-end
parser does not support, like generic type inheritance and type variables, or
when the analysis misidentifies an imported library alias.

For libraries that provide type hints in docstrings, which Joern does not parse,
we copy (but do not modify) the hints as type annotations ($<$100 LoC).
We manually copy dependencies into the analysis scope when discovered during
policy generation.
Because model loading policies are compositional, we
pre-compute policies for some frequently reused dependencies, including classes
from the Python standard library and PyTorch, and save them in a
class ``cache'' for \tool to access when it recognizes one of the classes in its
analysis.
Due to the complexity and prevalence of dynamically dispatched and C code
implementations in PyTorch, we supplement our analysis of PyTorch modules with
the \weightsonly policy in the class cache.

\subsection{RQ1: Malicious Model Blocking}\label{SEC:EVAL_MALICIOUS_BLOCKING}

\tool must protect loading programs from \pickle models with malicious payloads.
To evaluate this, we assess whether any of the generated \tool policies
allow malicious model executions.

\subsubsection{Methods}
For every harness program in our dataset, \tool generates a loading policy by
analyzing the library and associated model class.
We use \tool to enforce the generated policy while the harness attempts to load
all malicious models in our dataset.
For each malicious model, we consider the model blocked if \tool raises a
security exception during model loading or inference, preventing the payload
from executing.

Some library APIs only load the \pickle model after validating that the
accompanying model metadata is well-formatted
(\eg architecture, name, version).
For these libraries, we directly invoke \inlPython{pickle.loads} on the
malicious model, while enforcing the associated \tool policy.

\subsubsection{Results} \tolerance=1600 For all generated policies, \tool
\emph{prevents all (100\%) malicious models from executing their payloads}.
Since \tool's generated policies do not contain the dangerous callables
leveraged by the malicious models (\eg \texttt{eval()}, \texttt{system()}),
\tool's loader raises an exception for all malicious models.

\summarybox{
    \textbf{RQ1 Summary:} \tool generates policies that effectively
    prevent all malicious pickled models in our dataset from executing their
    payloads.
}

\subsection{RQ2: Benign Model Loading}\label{SEC:EVAL_SAFE_LOADING}

\begin{table*}[t!]
    \caption{
    \tool generates loading policies for popular libraries (see GitHub Stars),
    which are evaluated by loading popular models (see cumulative model
    downloads in March 2025).
    We compare the number callables observed in the models to the callables
    allowed by the policies, and the number of stub objects that are created or
    invoked when the policies exclude callables.
    We compare \tool's loading success rate with the \weightsonly.
    }
    \label{TABLE:static-policy-allow-benign-globals}
    \footnotesize
    \centering
    \setlength{\tabcolsep}{4pt} %
    \begin{tabular}{l|r r |r r r|r r r | r r r}
        \toprule
        & \multicolumn{2}{c|}{\textbf{Popularity}} & \multicolumn{3}{c|}{\textbf{Imports}} & \multicolumn{3}{c|}{\textbf{Invocations}} & \multicolumn{3}{c}{\textbf{Loading}}\\
        \textbf{Library} & Stars & Downloads & Observed & Allowed & Stub Objects  & Observed & Allowed & Stub Calls & \# Models & WOUp (\% ) & \tool (\%) \\
        \toprule
        CONCH~\cite{conch}                   &   \conchStars &                   \conchDownloads &                   \conchImportsObserved &                 \conchImportsAllowed &                  \conchImportsStubbed &                  \conchInvocationsObserved &                 \conchInvocationsAllowed &                  \conchInvocationsStubbed &                  \conchModelsTotal &                 \conchModelsWOU~(\conchWOUSuccessrate) &                          \conchModelsLoaded~(\conchModelsSuccessrate) \\
        FlagEmbedding~\cite{flagembedding}           &   \flagembeddingStars &           \flagembeddingDownloads &           \flagembeddingImportsObserved &         \flagembeddingImportsAllowed &          \flagembeddingImportsStubbed &          \flagembeddingInvocationsObserved &         \flagembeddingInvocationsAllowed &          \flagembeddingInvocationsStubbed &          \flagembeddingModelsTotal &         \flagembeddingModelsWOU~(\flagembeddingWOUSuccessrate) &                \flagembeddingModelsLoaded~(\flagembeddingModelsSuccessrate) \\
        flair~\cite{flair}                   &   \flairStars &                   \flairDownloads &                   \flairImportsObserved &                 \flairImportsAllowed &                  \flairImportsStubbed &                  \flairInvocationsObserved &                 \flairInvocationsAllowed &                  \flairInvocationsStubbed &                  \flairModelsTotal &                 \flairModelsWOU\phantom{00}~(\flairWOUSuccessrate) &                                \flairModelsLoaded\phantom{0}~(\flairModelsSuccessrate) \\
        GLiNER~\cite{gliner}                 &   \glinerStars &                  \glinerDownloads &                  \glinerImportsObserved &                \glinerImportsAllowed &                 \glinerImportsStubbed &                 \glinerInvocationsObserved &                \glinerInvocationsAllowed &                 \glinerInvocationsStubbed &                 \glinerModelsTotal &                \glinerModelsWOU~(\glinerWOUSuccessrate) &                              \glinerModelsLoaded~(\glinerModelsSuccessrate) \\
        huggingsound~\cite{huggingsound}            &   \huggingsoundStars &            \huggingsoundDownloads &            \huggingsoundImportsObserved &          \huggingsoundImportsAllowed &           \huggingsoundImportsStubbed &           \huggingsoundInvocationsObserved &          \huggingsoundInvocationsAllowed &           \huggingsoundInvocationsStubbed &           \huggingsoundModelsTotal &          \huggingsoundModelsWOU~(\huggingsoundWOUSuccessrate) &                  \huggingsoundModelsLoaded~(\huggingsoundModelsSuccessrate) \\
        LanguageBind~\cite{languagebind}            &   \languagebindStars &            \languagebindDownloads &            \languagebindImportsObserved &          \languagebindImportsAllowed &           \languagebindImportsStubbed &           \languagebindInvocationsObserved &          \languagebindInvocationsAllowed &           \languagebindInvocationsStubbed &           \languagebindModelsTotal &          \languagebindModelsWOU~(\languagebindWOUSuccessrate) &                  \languagebindModelsLoaded~(\languagebindModelsSuccessrate) \\
        MeloTTS~\cite{melotts}                 &   \melottsStars &                 \melottsDownloads &                 \melottsImportsObserved &               \melottsImportsAllowed &                \melottsImportsStubbed &                \melottsInvocationsObserved &               \melottsInvocationsAllowed &                \melottsInvocationsStubbed &                \melottsModelsTotal &               \melottsModelsWOU~(\melottsWOUSuccessrate) &                            \melottsModelsLoaded\phantom{0}~(\melottsModelsSuccessrate) \\
        Parrot\_Paraphraser~\cite{parrot}     &   \parrotStars &                  \parrotDownloads &                  \parrotImportsObserved &                \parrotImportsAllowed &                 \parrotImportsStubbed &                 \parrotInvocationsObserved &                \parrotInvocationsAllowed &                 \parrotInvocationsStubbed &                 \parrotModelsTotal &                \parrotModelsWOU~(\parrotWOUSuccessrate) &                              \parrotModelsLoaded~(\parrotModelsSuccessrate) \\
        PyAnnote~\cite{pyannote}                &   \pyannoteStars &               \pyannoteDownloads &               \pyannoteImportsObserved &             \pyannoteImportsAllowed &              \pyannoteImportsStubbed &              \pyannoteInvocationsObserved &             \pyannoteInvocationsAllowed &              \pyannoteInvocationsStubbed &              \pyannoteModelsTotal &             \pyannoteModelsWOU\phantom{00}~(\pyannoteWOUSuccessrate) &                    \pyannoteModelsLoaded\phantom{0}~(\pyannoteModelsSuccessrate) \\
        pysentimiento~\cite{pysentimiento}          &   \pysentimientoStars &         \pysentimientoDownloads &         \pysentimientoImportsObserved &       \pysentimientoImportsAllowed &        \pysentimientoImportsStubbed &        \pysentimientoInvocationsObserved &       \pysentimientoInvocationsAllowed &        \pysentimientoInvocationsStubbed &        \pysentimientoModelsTotal &       \pysentimientoModelsWOU~(\pysentimientoWOUSuccessrate) &              \pysentimientoModelsLoaded~(\pysentimientoModelsSuccessrate) \\
        sentence\_transformers~\cite{sentence-tran}  &    \sentencetransformersStars &    \sentencetransformersDownloads &    \sentencetransformersImportsObserved &  \sentencetransformersImportsAllowed &   \sentencetransformersImportsStubbed &   \sentencetransformersInvocationsObserved &  \sentencetransformersInvocationsAllowed &   \sentencetransformersInvocationsStubbed &   \sentencetransformersModelsTotal &  \sentencetransformersModelsWOU~(\sentencetransformersWOUSuccessrate) &  \sentencetransformersModelsLoaded~(\sentencetransformersModelsSuccessrate) \\
        super-image~\cite{super-image}             &   \superimageStars &             \superimageDownloads &             \superimageImportsObserved &           \superimageImportsAllowed &            \superimageImportsStubbed &            \superimageInvocationsObserved &           \superimageInvocationsAllowed &            \superimageInvocationsStubbed &            \superimageModelsTotal &           \superimageModelsWOU~(\superimageWOUSuccessrate) &                  \superimageModelsLoaded~(\superimageModelsSuccessrate) \\
        TNER~\cite{tner}                    &   \tnerStars &                   \tnerDownloads &                   \tnerImportsObserved &                 \tnerImportsAllowed &                  \tnerImportsStubbed &                  \tnerInvocationsObserved &                 \tnerInvocationsAllowed &                  \tnerInvocationsStubbed &                  \tnerModelsTotal &                 \tnerModelsWOU~(\tnerWOUSuccessrate) &                                \tnerModelsLoaded~(\tnerModelsSuccessrate) \\
        tweetnlp~\cite{tweetnlp}               &   \tweetnlpStars &               \tweetnlpDownloads &               \tweetnlpImportsObserved &             \tweetnlpImportsAllowed &              \tweetnlpImportsStubbed &              \tweetnlpInvocationsObserved &             \tweetnlpInvocationsAllowed &              \tweetnlpInvocationsStubbed &              \tweetnlpModelsTotal &             \tweetnlpModelsWOU~(\tweetnlpWOUSuccessrate) &                    \tweetnlpModelsLoaded~(\tweetnlpModelsSuccessrate) \\
        YOLOv5~\cite{yolov5}                 &   \yolovfiveStars &              \yolovfiveDownloads &              \yolovfiveImportsObserved &            \yolovfiveImportsAllowed &             \yolovfiveImportsStubbed &             \yolovfiveInvocationsObserved &            \yolovfiveInvocationsAllowed &             \yolovfiveInvocationsStubbed &             \yolovfiveModelsTotal &            \yolovfiveModelsWOU\phantom{00}~(\yolovfiveWOUSuccessrate) &                  \yolovfiveModelsLoaded\phantom{0}~(\yolovfiveModelsSuccessrate) \\
        YOLOv11 (ultralytics)~\cite{yolov11}   &   \yolovelevenStars &            \yolovelevenDownloads &            \yolovelevenImportsObserved &          \yolovelevenImportsAllowed &           \yolovelevenImportsStubbed &           \yolovelevenInvocationsObserved &          \yolovelevenInvocationsAllowed &           \yolovelevenInvocationsStubbed &           \yolovelevenModelsTotal &          \yolovelevenModelsWOU\phantom{00}~(\yolovelevenWOUSuccessrate) &                  \yolovelevenModelsLoaded\phantom{0}~(\yolovelevenModelsSuccessrate) \\
        \midrule
        \textbf{Total} & & & &  &  &  &  &  & \modelsTotal & \modelsTotalWOULoaded~(\modelsTotalWOUSuccessrate) & \modelsTotalPickleballLoaded~(\modelsTotalSuccessRate) \\
        \textbf{Average} & & & &  &  &  &  &  & & \modelsAvgWOU & \modelsAvgPickleball \\
        \bottomrule
    \end{tabular}
\end{table*}

\tool must let users load and perform tasks with benign models.
Its policies must not be so restrictive that the models are unusable.
We therefore evaluate \tool's policies for loading and correctly using the
benign models in our dataset
(\S\ref{SEC:EVAL_BENIGN_METHODS} and \S\ref{SEC:EVAL_BENIGN_RESULTS}).
To ensure robustness of the loaded model despite lazy enforcement, we further
test the successfully loaded models that contain stub objects
(\S\ref{SEC:EVAL_BENIGN_ROBUST_METHODS} and
\S\ref{SEC:EVAL_BENIGN_ROBUST_RESULTS}).

\subsubsection{Methods}\label{SEC:EVAL_BENIGN_METHODS}
We measure \tool's ability to generate and enforce policies that correctly
load and execute benign models.
For each library in our dataset, we generate a model loading policy.
Then, we enforce the policy while using the library's test harness to load each
library's models.
Once loaded, we test the model by performing one inference task with a test
input, and capture the output.
For comparison, we
then re-execute the test harness \textit{without} enforcing any policy (by
using the regular \pickle module), and capture the inference result.
We consider the model load a success when (1) \tool loads the model without
raising exceptions, and (2) the inference results are equivalent between the
\tool and unrestricted environments.

\subsubsection{Results}\label{SEC:EVAL_BENIGN_RESULTS}
\tool generates policies that, when enforced, correctly load and execute
\modelsTotalSuccessRate~of benign models in the dataset
(\cref{TABLE:static-policy-allow-benign-globals}).
In most cases, the policies contain all callables
(\cref{TABLE:static-policy-allow-benign-globals} -- Imports and Invocations
Allowed) that are seen in the model traces (Imports and Invocations Observed).
In some cases, \tool's policies do not include callables that are included in
the models (flair, PyAnnote, YOLOv5, and YOLOv11), resulting in the creation of
a stub object that is occasionally invoked, hence leading to a security
violation.
We investigated the failed models/\-cases to determine their causes:
\begin{itemize}
    \item \textbf{Attributes set after initialization}: \tool fails to
        identify attribute types that are set outside of the type declaration.
        For example, after initializing the model object, some libraries allow
        users to write training metadata to the model, including data for the
        optimizer used and paths to output files.
        In many cases (\eg~12 flair models), \tool misses callable types set
        this way but still successfully loads the model, since this metadata is
        not used.
        However, for one flair model and three PyAnnote models, a metadata
        object is invoked during loading.

    \item \textbf{Follow-on \pickle loading}: \tool fails to load two models
        from the MeloTTS library after they have been loaded, due to additional
        \pickle loading during inference.
        \tool's policy includes all callables needed to load the MeloTTS models.
        However, during inference, an additional \pickle model is loaded that
            invokes a disallowed callable, resulting in a security
            violation.

    \item \textbf{Library version drift}: one PyAnnote model fails to load for
        legacy reasons: it uses a callable that was included in models created
        with previous versions of PyAnnote.
        The callable's class declaration exists in the library code base as an
        unused stub, with a comment that it is needed for backward compatibility
        reasons, but is otherwise unused.
        Therefore, \tool's analysis failed to recognize it as necessary for
        model loading.

    \item \textbf{Namespace inconsistency}: the remaining \texttt{YOLOv5} and
        \texttt{YOLOv11} models use inconsistent naming conventions.
        For example, a policy includes the callable
        \scode{yolov5.models.common.Conv}; however, the model refers to this
        callable as \scode{models.common.Conv}, while referring to other
        callables by the full \scode{yolov5.*} namespace.
\end{itemize}

\subsubsection{Lazy Enforcement Robustness -- Methods}\label{SEC:EVAL_BENIGN_ROBUST_METHODS}
Due to \tool's lazy enforcement, models can load successfully with incomplete
attributes.
To further ensure that the benign models are robustly instantiated with all
attributes needed for inference, we evaluate these models with a more rigorous
test suite of inputs.
We investigated the libraries that successfully load models with stub objects,
\ie Flair, PyAnnote, YOLOv5, and YOLOv11.

For each library, we find an extensive evaluation dataset to test each loaded
model:
    for Flair, we used various Named Entity Recognition and Universal
    Dependencies~\cite{universaldependencies} datasets that come pre-packaged with
    the Flair library~\cite{flair-datasets};
    for PyAnnote, we used the AISHELL-4 speech dataset~\cite{aishell4};
    for YOLOv5 and YOLOv11, we used the 2017 Test Images Common Objects in Context
    dataset~\cite{microsoft-coco}.

For each model that \tool successfully loads, we evaluate the model on the
dataset and ensure that the models do not raise security errors (\ie they do not
access any stub objects).

\subsubsection{Lazy Enforcement Robustness --
Results}\label{SEC:EVAL_BENIGN_ROBUST_RESULTS}
All models yield the same results when loaded with \tool, compared to when
loaded with the regular, unrestricted \picklemachine.
None of the models raise security errors during dataset evaluation, indicating
that models are correctly instantiated, despite using stub objects.

\summarybox{\textbf{RQ2 Summary:} \tool generates policies that safely load and execute
\modelsTotalSuccessRate~of benign pickled models in our dataset.
}

\subsection{RQ3: Performance}\label{SEC:EVAL_PERFORMANCE}

\tool must be fast enough for practical use in developer and user tasks.
We analyze two aspects of \tool's performance:
  (1) the time to generate a policy for a class, which is an offline, one-time
  analysis cost (\S\ref{SEC:OVERVIEW_STATIC}),
  and
  (2) the additional runtime overhead of enforcing a policy to load and use a
  model, compared to the regular \picklemachine.

\subsubsection{Methods}
To measure the time to generate policies, we execute \tool's policy generator
three times for each library in our dataset and compute the average between the
three.
We measure the real time using the Python \scode{time} library.
We run this experiment on a laptop with a 14-core Intel~i7~CPU and 32GB of RAM,
representing a commodity developer environment.

To measure the additional runtime overhead of \tool's policy enforcer, we
isolate and record the time each harness program spends invoking the \pickle~
\inlPython{load} function during model loading.
We first execute harness program with the unrestricted \picklemachine
environment to load a benign model.
Then, we perform the same execution with \tool enabled.
For fair comparison, we ensure that the unrestricted environment always uses
the Python implementation of the \picklemachine, instead of an optimized C
implementation.
We run this experiment on a server with a 32-core AMD~EPYC~7502 processor and
256GB of RAM (Ubuntu 24.04); this is used for the attached hard-drive
space for interacting with the hundreds of models in our dataset.

\subsubsection{Results}
\begin{figure}[t]

    \centering
    \includegraphics[width=\columnwidth]{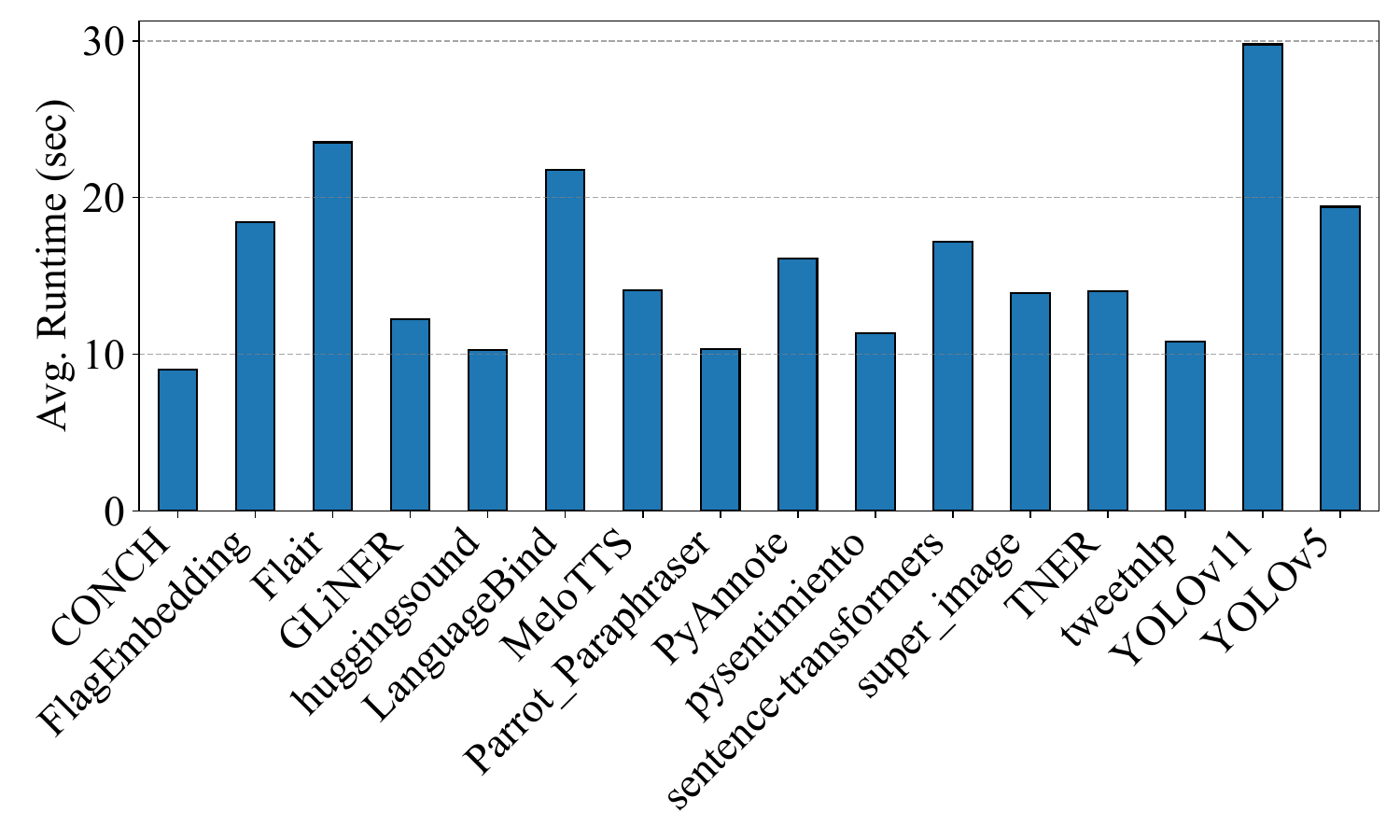}
        \caption{
    Time to generate a policy for each library class in dataset (averaged over 3 runs).
    This is a one-time step that can be integrated into existing workflows ---
      either by library maintainers in the library's release process,
      or
      by a user, prior to loading the model.
    }
    \label{FIG:POLICY_GENERATION_PERFORMANCE}
\end{figure}

\begin{figure}[t]
    \centering
    \includegraphics[width=\columnwidth]{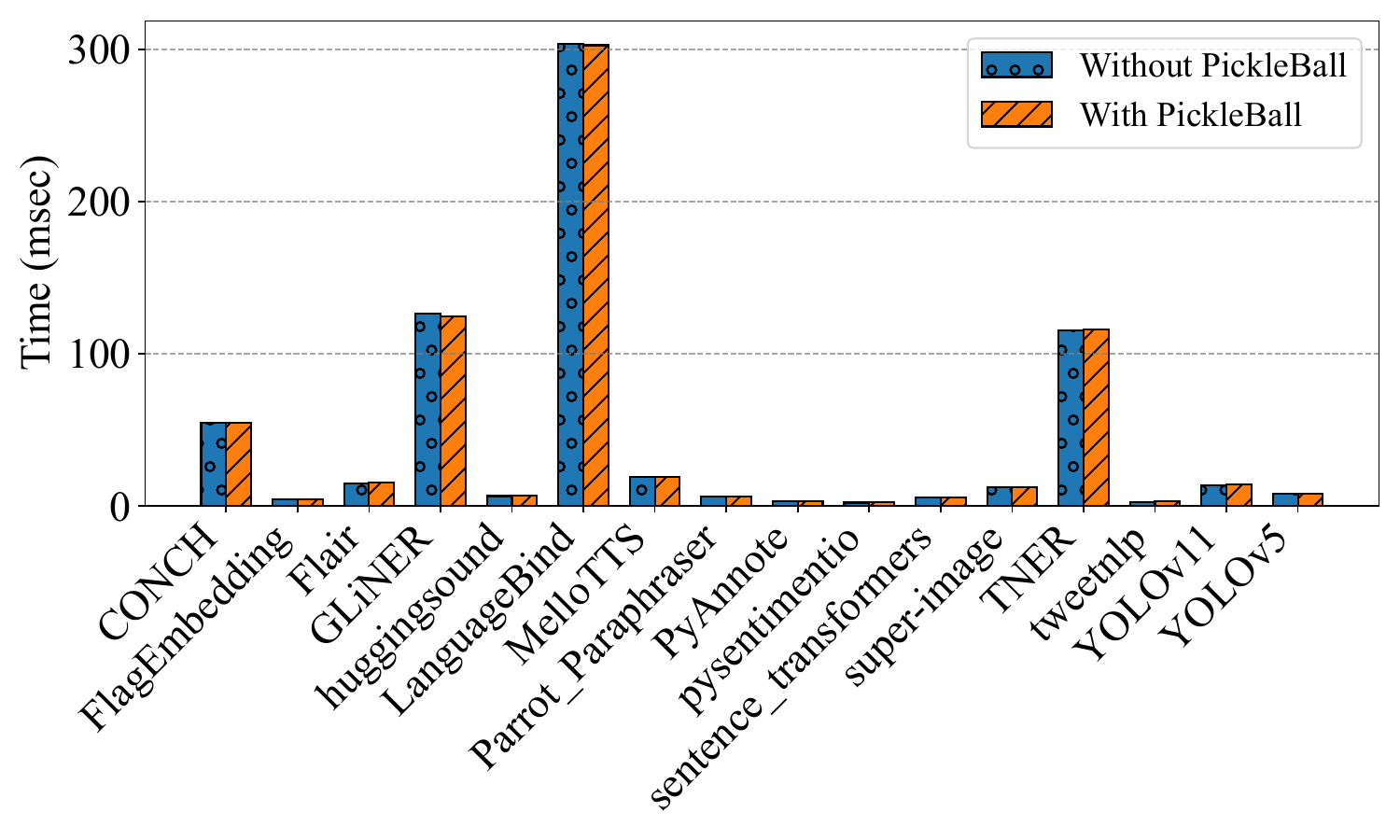}

    \caption{
    Time spent executing \scode{pickle.load} in test loading program, with and
    without \tool (averaged over 3 runs after 1 warmup run).
    \tool incurs a median runtime overhead of \runtimeEnforcementOverheadPer~and
    average runtime overhead of \runtimeEnforcementAvgOverheadPer.
    \andreas{This figure really should be one that shows \% increase.}
        }
    \label{FIG:ENFORCEMENT_PERFORMANCE}
\end{figure}

\tool generates policies for all libraries in a median of
\timeGenerationMedian~seconds, with minimum \timeGenerationMinValueSec seconds
(\timeGenerationMinLibrary library) and maximum
\timeGenerationMaxValueSec~seconds (\timeGenerationMaxLibrary library)
(Figure~\ref{FIG:POLICY_GENERATION_PERFORMANCE}).
This policy generation execution time is reasonable for integration within
project build systems, as the policy needs only to be generated when the
analyzed library source code is modified.
\tool's policy enforcer incurs negligible overhead, with a
\runtimeEnforcementOverheadMs (\runtimeEnforcementOverheadPer)
median runtime overhead compared to the unrestricted \picklemachine, as
depicted in Figure~\ref{FIG:ENFORCEMENT_PERFORMANCE}.

\summarybox{\textbf{RQ3 Summary:} \tool policies are generated in a median
\timeGenerationMedian~seconds across the evaluation libraries, \tool incurs a
median runtime overhead of \runtimeEnforcementOverheadMs when loading models.
}

\subsection{RQ4: Comparison to SOTA}\label{SEC:EVAL_COMPARISON_TO_SOTA}

\begin{table}[t!]
\centering

    \caption{Comparison of \tool to SOTA alternatives.
    Model scanning tools achieve low false positives on our dataset, but
    misclassify malicious models.
    Restricted loaders (including \tool) are secure, at the cost of blocking
    benign models.
    \tool loads more benign models than the \weightsonly due to its custom
    policies for each model class.
    }
    \label{TABLE:sota-comparison}
    \resizebox{\columnwidth}{!}
    {
    \begin{tabular}{l c c c c c c}
        \toprule
        \textbf{Tool} & \textbf{\# TP} & \textbf{\# TN} & \textbf{\# FP} & \textbf{\# FN} & \textbf{FPR} & \textbf{FNR} \\
        \toprule
        \modelscan~\cite{protectai_modelscan} & 75 & 236 & 16 & 9 & 6.3\% & 10.7\% \\
        \modeltracer~\cite{casey2024largescaleexploitinstrumentationstudy} & 44 & 252 & 0 & 40 & 0\% & 47.6\% \\
        \\
        Weights-Only Unpickler~\cite{weightsOnlyUnpickler}& \nmaliciousdatasetmodels & \modelsTotalWOULoaded & \modelsTotalWOUFailed & 0 & \modelsTotalWOUFalureRate & 0\% \\
        \tool (our work) & \nmaliciousdatasetmodels & \modelsTotalPickleballLoaded & \modelsTotalPickleballFailed & 0 & \modelsTotalFailureRate & 0\% \\
        \bottomrule
    \end{tabular}
    }
\end{table}

We compare \tool with three existing state-of-the-art (SOTA) tools that share
the same goal of defending against our threat model described in
Section~\ref{SEC:MOTIVATION_THREAT_MODEL}.
As discussed in Section~\ref{SEC:BACKGROUND_PICKLE}, existing \pickle model
defense tools fall into two categories: model scanners and restricted loading
environments (like \tool).

We compare against two model scanners: \modelscan~\cite{protectai_modelscan},
and the scanner implemented by
Casey~\etal~\cite{casey2024largescaleexploitinstrumentationstudy} (henceforth
\modeltracer).\!%
    \footnote{The authors provided access to the tool for evaluation
    purposes.}
\modelscan is a static analysis tool that applies a denylist
to make determinations about models, and is integrated into Hugging Face.
\modeltracer is a dynamic analysis tool that traces the model's invocations
while it is loaded via \texttt{pickle.loads()}/\-\texttt{torch.load()},
and similarly applies a denylist.

We compare against one  restricted loading environment: the \weightsonly~\cite{weightsOnlyUnpickler}. The \weightsonly loads models by only permitting them access to callables in a rigid (but manually configurable) allowlist policy.

\subsubsection{Methods}
We evaluate the model scanning tools by providing each model in our dataset as
an input to the tool.
We expect the model scanners to alert when provided a malicious model input, and
otherwise not to alert.

We consider correctly identified malicious models as true positives, correctly
identified benign models as true negatives, incorrectly identified malicious
models as false negatives, and incorrectly identified benign models as false
positives.

We evaluate the \weightsonly by attempting to load each model in our dataset
using the PyTorch loading API with the \weightsonly enabled.
We use the \weightsonly's default policy while loading models.
We expect restricted loading environments like the \weightsonly and \tool to
succeed when loading benign models and to raise exceptions when loading
malicious models.

For parity when comparing with the model scanning tools, we consider raising an
exception during malicious model loading as a true positive, correctly loading a
benign model as a true negative, incorrectly rejecting a benign model as a false
positive, and incorrectly loading a malicious model as a false negative.

\vspace{-5pt}

\subsubsection{Results} Comparisons of the tools are shown in
\cref{TABLE:sota-comparison}.
The model scanning tools resulted in few (16) false positives, while the
restricted loaders resulted in 0 false negatives.
\cref{TABLE:static-policy-allow-benign-globals} compares the success rate of
\tool and the \weightsonly when loading benign models.

\modelscan incorrectly identified 9 malicious models as benign (false
negatives) and did not report false positives. We identified three categories
of \modelscan's false negatives: (1) five models implement payloads using
callables that are not included in \modelscan's rigid denylist; (2) three
models use dynamic runtime operations (\eg \inlPython{numpy.load()}) to load
additional payloads that \modelscan fails to statically identify; and (3) one
model uses multiple \texttt{STOP} \pickle opcodes, resulting in \modelscan
terminating its analysis after reaching the first one and missing the rest of
the malicious payload. While this last model would not execute its malicious
payload when executed by the \picklemachine, it could be loaded in non-standard
ways by another malicious \pickle program to execute its payload. Categories
(1) and (2) are fundamental limitations to using a static analysis denylist
approach: the denylist cannot be complete and can be subverted.

\modeltracer successfully identified 44 malicious models but missed the
remaining 40, resulting in a high false negative rate of 47.6\%, but did not
report false any positives.
\modeltracer's false negatives appear from its limited denylist: it alerts on
models that invoke the \texttt{execve}, \texttt{connect}, \texttt{socket}, or
\texttt{chmod} system calls.
\modeltracer does not consider file access operations to be indicators of
malicious behavior, so malicious models that perform dangerous file reads and
writes are not identified.
This once again highlights the scanning limitation of relying on an incomplete
denylist to indicate malicious behavior.

The \weightsonly, like \tool, prevents all malicious models from loading.
However, it incorrectly blocks \modelsTotalWOUFailed~benign models from loading,
compared to
\tool's \modelsTotalPickleballFailed.
However, \tool's custom generated policies load additional models that the
\weightsonly can not.

\summarybox{
    \textbf{RQ4 Summary:} While \tool prevents all malicious models from
    loading, model scanning tools fail to identify all malicious models. The
    \weightsonly is also effective at preventing malicious models from loading,
    but is less effective than \tool at loading benign models due to its limited
    default policy.
}

\section{Discussion and Future Work} \label{SEC:DISCUSSION}

\para{\tool's remaining attack surface}
\tool reduces the available attack surface by significantly restricting access
to callables, but it does not guarantee that the remaining callables cannot be
composed in a malicious payload (\S\ref{SEC:OVERVIEW_GUARANTEES}).
\tool could be compromised by code reuse
techniques~\cite{park2022fugio,dahse2014code} to stitch together permitted calls
to construct an exploit.
Thus far, no such attacks have been observed, but the question remains: can we
generate malicious payloads that obey the policy constraints enforced by \tool
and the \weightsonly?

Huang et al.~\cite{huang2022painpickle} studied manual implementations of
restricted unpicklers in the general (non-ML) setting, and devised attack
strategies to overcome callable allowlists.
When faced with well-implemented restricted unpicklers (\ie when recursive
attribute look-ups and indexing are disallowed by design, as in \tool and the
\weightsonly), their approach degrades to manual policy inspection.
Liu et al.~\cite{liu:2025:artofhideandseek} implemented \textsc{PickleCloak} to
automatically detect useful \pickle gadgets, but apply it to bypass scanner
deny-lists rather than restricted loader allow-lists.
Future work should identify properties of callables to automatically distinguish
whether a callable can be used maliciously to bypass restricted allow-lists.

\para{Long-term outlook for \pickle in ML}
\Pickle remains a popular model format (\S\ref{SEC:MOTIVATION}), despite more
secure alternatives.
Each major model format provides tradeoffs in flexibility (\pickle), security
(SafeTensors), and efficiency (GGUF).
\Pickle is flexible, as it can serialize virtually any Python object, including
complex models and non-standard data structures.
SafeTensors was developed for security-sensitive deployments, with a
structured, memory-mapped format.
GGUF maximizes performance on inference-optimized runtimes.
\textit{Because these formats are complementary and used as defaults in
different popular frameworks and ecosystems, we expect them to coexist going
forward.}
\tool does not discourage the adoption of secure alternatives to \pickle, but
provides a secure option for the large and growing \pickle
population.

One rapidly evolving area of ML models is large language models (LLMs), where
\pickle still appears despite major industry leaders releasing foundation models
in the SafeTensors and GGUF formats.
Popular foundation models like LLaMA-4~\cite{llama4_scout},
Qwen-3~\cite{qwen3_06B}, and
Deepseek-R1~\cite{deepseek_r1_distill_qwen_32B} are
encouragingly released with the SafeTensors format (although some, like
LLaMA-3.1~\cite{llama_31_pickle}, still provide a \pickle model backup).
However, foundation models are often adapted (\eg~fine-tuned) and redistributed,
often with new formats and artifacts, including \pickle.
For example, we observed instances of models that are fine-tuned from
LLama-4~\cite{madilcy_arabic_medical_llama4,ilya_bs1_llama4_scout_interpreter_model},
Qwen-3~\cite{TarhanE_Qwen3,sarahbadr_mnlp_m2_dpo_model}, and
Deepseek-R1~\cite{tttx_models-p10-ttt-18feb-fixed-sft-clip-step1} and
distributed with an additional \pickle file that represents the training
arguments used during fine-tuning.
Even when secure formats are adopted for foundation LLM models, \pickle
continues to persist in the LLM ecosystem, which is consistent with our
analysis (\S\ref{SEC:MOTIVATION_SURVEY}) and justifies the need for
\tool.

\para{Generalizing \tool}
\tool is designed to protect \pickle model loading and is evaluated on models
found on Hugging Face, but its approach can generalize to protect
  (1)~other model formats;
  and
  (2)~other \pickle applications.
\tool aims to protect \pickle deserialization for ML models, but its approach
does not rely on any ML-specific properties.

This approach allows \tool to protect other applications that receive
\pickle data.
To generalize, \tool's analysis requires that \textit{the intended type of the
\pickle object is known before loading}~(\S\ref{SEC:OVERVIEW_STATIC}).
In the ML setting, this is reasonable because the protected program is a
\textit{client application}.
Other approaches are needed when the security analysis does not know
\textit{a priori} the intended type of the serialized object%
~\cite{david_quack_2024,zhang:2024:javadeserializationdefense}.

\tool's approach works for the \pickle format because it
  has (dangerously) expressive deserialization operations, and
  is used by trusted libraries that implement their own custom model
  classes. %
Other model formats that meet these criteria are candidates for protection in
the \tool approach.
Zhu et al. show that the TensorFlow SavedModel format has undesirable
operations~\cite{zhu:2025:mymodelismalwaretoyou};
  libraries that extend the TensorFlow model class with their own custom
  behaviors could use a \tool approach to restrict the allowed behaviors, but
  we are not aware of any that do.
Formats like SafeTensors and GGUF are not known to have dangerous operations;
if any were discovered, then a \tool approach might apply for identifying when
to permit certain operations.
We aim to explore which other model formats meet these criteria for \tool to
assist in securely loading.

We evaluated \tool using models sourced from Hugging Face, but \tool will work
similarly for models from any platform.
The inputs to \tool are ML libraries and \pickle models, which are ML artifacts
that are not tied to the hosting platform.
Hugging Face is the largest model hosting platform, with over 1.8M models
available in July 2025.
ModelScope~\cite{modelscopehub} is a recent hub managed
by China's Alibaba and hosts 80K models.
It imitates Hugging Face's design and likewise has models with varying
serialization formats.
Other model communities, including Qualcomm AI Hub~\cite{qualcommaihub},
PyTorch Hub~\cite{pytorchhub}, and TensorFlow
Hub~\cite{tensorflowhub}, have fewer than 500 models each and many are also
hosted on Hugging Face.
The ONNX Model Zoo~\cite{onnxmodelzoo} is now deprecated and archived on Hugging
Face.
Hugging Face models are representative of the kinds of models that \tool
defends against.

\para{Policy maintenance and distribution}
\tool's intended workflow is that when a library is updated, its \tool policy
would be updated as well.
\tool makes policy maintenance easy for users with fast policy generation,
incremental changes, and opportunities for seamless distribution.

\tool generates policies quickly, completing in under 30 seconds for each
library in our evaluation (\S\ref{SEC:EVAL_PERFORMANCE}, \cref{FIG:POLICY_GENERATION_PERFORMANCE}).
This is reasonable for a task that must only occur when the library changes,
and not every time \tool loads a model.

In practice, library updates result in either incremental policy changes
or clearly documented breaks in supported model versions, leading to easier
policy maintenance.
After our evaluation concluded, we noticed one library, \texttt{FlagEmbedding}, receive
updates (commit \texttt{bf6b649} to \texttt{875fd4f}), but \tool
produced policies before and after with a 90\% Jaccard similarity index, and
which successfully loaded the same models in our dataset.
When the library model class changes significantly, the library cannot load
existing models, but we find it easy in practice to match models with a
supported library version, due to the model documentation, as we do in our
evaluation (\S\ref{SEC:EVAL_BENIGN_RESULTS}).
For example, we easily distinguish all models belonging to \texttt{YOLOv5} and its newer
version, \texttt{ultralytics}.

\tool provides opportunities for easier policy maintenance if it gets adopted
further in the model development life cycle.
When library maintainers adopt \tool to generate policies automatically in the
library release process, they can provide the updated policies alongside
the libraries directly to users.

\para{Removing library pre-processing}
To account for implementation limitations (\S\ref{SEC:IMPLEMENTATION}) when
evaluating the fundamental \tool idea, we perform manual pre-processing of some
libraries (\S\ref{SEC:EVAL_DATASET}), but future engineering work will remove
this step.
The purpose of the pre-processing is to overcome implementation limitations of the
underlying static analysis framework that \tool depends on to produce an
accurate type-annotated AST.
To account for these limitations, we apply the following pre-processing steps:
\begin{itemize}
\item Remove generic type syntax from class inheritance statements.
\item Copy type hints in comments into type annotation form.
\item Copy relevant dependencies into the analysis scope.
\item Reference the \weightsonly policy when a class inherits the \texttt{torch.nn.Module} class.
\end{itemize}
Manual source code modifications are applied to five out of \ndatasetlibraries
libraries in our dataset, and each account for between $\sim$10 and $\sim$100
modified LoC.
Future engineering work to improve the underlying static analysis framework will
remove the need for manual source code pre-processing.

\section{Related Work} \label{SEC:RELATED}

\para{ML Model Loading Security}
\Pickle is not intended for untrusted data, but its proliferation as a model
format created a security problem.
To bring attention to the issue, security company Trail of Bits released the
\scode{fickling} tool for manipulating and analyzing \pickle programs in ML
models~\cite{fickling}.
The fickling module provides manually-crafted allow-lists for select ML
libraries, similarly to \tool, but requires expert-led audits to maintain the
allow-lists, where \tool aims for automatic policy creating.
New proposals for identifying malicious \pickle models include dynamic
\cite{casey2024largescaleexploitinstrumentationstudy} and static
\cite{zhao2024MalHug} scanners.
Scanners take a
model as input, and attempt to make an assessment of it based on fixed rules
about malicious behaviors.
Instead, \tool takes a model and the source code
library that allegedly produced the model for context, and produces policies for
the model based on that context.
We showed in
\cref{SEC:EVAL_COMPARISON_TO_SOTA} that \tool's tailor-made policies result in
no false negatives while comparative model scanners do
produce false negatives due to their fixed rules.

\para{Deserialization Attacks and Defenses}
Deserialization vulnerabilities exist beyond the ML context.
PainPickle~\cite{huang2022painpickle} explored Python \pickle security by
creating a taxonomy of errors in custom Unpickler implementations, and devised
attack strategies.
We use their contributions (and suggestions) to guide the proper implementation of \tool's loader.

Other programming languages also have insecure deserialization APIs that
need to be secured.
Quack~\cite{david_quack_2024} proposes a
generic deserialization defenses for PHP by employing a ``static duck typing''
static analysis, and Zhang et al.~\cite{zhang:2024:javadeserializationdefense} propose a
static analysis defense for Java.
Python's \pickle deserialization API is more expressive than those in PHP
or Java, which are unable to directly invoke functions; this expressivity
adds complexity to security policies.

\para{Querying Code-graphs for Software Security}
Graph representations are well-established for general program analysis tasks.
For security specific tasks, Joern~\cite{yamaguchi2014joern} introduced the Code
Property Graph (CPG), a data structure that combines classic program analysis
concepts into a representation that is easy queried to identify vulnerabilities.
Follow on work
ODGEN~\cite{odgen} extended Joern's
CPGs into an Object Dependence Graph (ODG), capturing interactions from the
object's point of view to detect vulnerabilities in \NJS packages.
RogueOne~\cite{sofaer_rogueone_2024} further evolved ODGs to form a data-flow
relationship graph, fully capturing data-flows among objects.
QL~\cite{moor:2007:QL} and Datalog~\cite{schafer:2010:typeinferencedatalog}
based approaches inspired the CodeQL~\cite{codeQLforsecurity} query platform,
which is used for vulnerability variant analysis tasks.
\tool's implementation uses these code querying features and extends them to
improve the accuracy of the program types recovered. Improvements to these
program representations can lead to improved accuracy of \tool's policies.

\section{Conclusion}

Serialization and deserialization enable code and data exchange.
Many recent works observe security vulnerabilities in deserialization,
across various programming languages and contexts.
We specifically examined the security vulnerabilities in pre-trained model
deserialization that result from the use of (dynamically typed) Python and the
reliance on Python's (insecure) \pickle format.
We found that
  \pickle is common among the most popular models on Hugging Face,
  and
  that existing defenses are insecure or inapplicable to a substantial fraction of these models.
Our \tool approach applies a novel program analysis to add greater security
to model deserialization.
In our evaluation, we demonstrated that \tool supported most existing benign models while preventing
all known attacks in malicious models.
We believe \tool is a promising complement to existing security resources in the
pre-trained model ecosystem.%

\section*{Research Ethics}
We considered the stakeholders involved in this study~\cite{davis2025guide}
  and believe our results offer a net benefit.
Our primary contribution is \tool, a defense that improves security for ML models.
Our Hugging Face study abides by the platform policies for API use.
We identified some shortcomings of existing defenses, and we
followed a responsible disclosure process to report them. %

\section*{Data Availability}

The \tool source code is maintained at:
\url{https://github.com/columbia/pickleball}.
Our complete evaluation artifact is made available at:
\url{https://doi.org/10.5281/zenodo.16974644}.
The artifact contains (1) \tool source code, (2) a dataset of malicious
and benign models, and (3) survey data of the \pickle model ecosystem.

\section*{Acknowledgements}
We thank our anonymous reviewers for their valuable feedback, and
Joseph Lucas, John Irwin, and Suha Hussain for their insightful
discussions of \pickle security.
We thank Beatrice Casey for providing access to the \textsc{ModelTracer}
tool~\cite{casey2024largescaleexploitinstrumentationstudy}, and to Shenao Wang
for sharing malicious models discovered on Hugging
Face~\cite{zhao2024MalHug}.
This work is supported by the National Science Foundation (NSF), through award CNS-2238467.
Andreas Kellas is supported by the Department of Defense (DoD) National Defense
Science and Engineering (NDSEG) Fellowship program.
Yaniv David is supported by the Center for Computer Engineering at the Technion.
James Davis is supported by funding from Socket.
Junfeng Yang is supported by funding from Google.

\bibliographystyle{ACM-Reference-Format}
\bibliography{main}

\ifextended
\clearpage

\appendix

\section*{Outline of Appendices}

\noindent
The appendix contains the following material:

\begin{itemize}[leftmargin=12pt, rightmargin=5pt]

\item \cref{HFMeasureAPdix}: Hugging Face Measurement Supplemental Details.
\item \cref{SEC:MOTIVATION_SCANNERS}: Techniques for Bypassing Model Scanners.
\item \cref{ImpAPdix}: \tool Implementation Details.
\item \cref{SEC:APDX:DATASET}: Libraries and Models in Benign Dataset.

\end{itemize}

\section{Hugging Face Measurement} \label{HFMeasureAPdix}

Our measurement of model serialization formats in Hugging Face used file extensions to determine model serialization format.
The mapping from extension to format assignment is as follows:

\begin{itemize}
    \item \textit{\pickle}: We used the core \pickle file extension and those of variants and wrappers (\texttt{.pkl}, \texttt{.pickle}, \texttt{.joblib}, and \texttt{.dill}). We also used those of PyTorch-specific variants: \texttt{.pt}, \texttt{.pth}, and \texttt{.bin}.
    \item \textit{SafeTensor}: We used the \texttt{.safetensor} extension.
    \item \textit{GUFF}: We used the \texttt{.gguf} extension.
    \item \textit{Other or missing}: We included additional extensions often associated with serialized model data, such as \texttt{.h5}, \texttt{.hdf5}, \texttt{.ckpt}, \texttt{.model}, \texttt{.pb}, \texttt{.npy}, \texttt{.npz}, \texttt{.onnx}, \texttt{.msgpack}, \texttt{.nemo}, \texttt{.wav}, and \texttt{.keras}.
    Some models do not include actual model files; instead, their model cards provide instructions for loading the model from a third-party library. In our study, we marked these models as ``missing files''.

\end{itemize}

When models included files with multiple extensions, we categorized them as having multiple file types available.
Our analysis focuses on the package level within Hugging Face, \ie on the most recent version of each package. While some packages may have multiple versions, the package name is the primary indicator of a new package, as noted in prior work \cite{jiang2023ptmreuse, jiang2024naming}.

We acknowledge that file extensions serve as a proxy for the file’s actual content. While they suggest the type of content, they do not attest it.

\para{``Unusual'' Callables}
We analyzed \tracedRepositories repositories containing pickled models, and
identified \tracedRepositoriesDisallowed repositories containing models with
``unusual'' callables, \ie those that are rejected during loading by the
\weightsonly. We show the set of observed ``unusual'' callables in
\cref{FIG:UNUSUAL_CALLABLES}.

\section{Bypassing Model Scanners}\label{SEC:MOTIVATION_SCANNERS}

Model scanners use a denylist approach to determine whether a model is
malicious, but \pickle denylists are circumventable, which we demonstrate by
example:
we construct ``malicious'' \pickle programs that state-of-the-art scanners
misclassify as benign.

We identify two strategies for bypassing model denylists to guide the implementation
of the models:
\begin{enumerate}
\item invoking callables that are \textit{missed by the denylist}; and
\item \textit{indirectly} importing and invoking disallowed callables.
\end{enumerate}
We demonstrate the efficacy of each strategy with constructed \pickle programs
that bypass both static (\ie~\modelscan~\cite{protectai_modelscan}) and dynamic
(\ie~\modeltracer~\cite{casey2024largescaleexploitinstrumentationstudy})
scanners, and disclose the bypasses to the scanner authors.

To demonstrate (1), we constructed a malicious pickle program that writes to an
attacker-chosen file using APIs from the \texttt{pathlib}~\cite{python-pathlib}
Python library, rather than the builtin \texttt{open} and \texttt{write}
functions.
\modelscan fails to recognize this as malicious because the \texttt{pathlib}
APIs are not part of its denylist~\cite{modelscan-denylist};
\modeltracer fails because it does not classify file write operations as
malicious.

To demonstrate (2), we constructed a \pickle program that invokes the
\texttt{os.system} function (which is \emph{disallowed} by the \modelscan
scanner), by importing it through an allowed module (in our case,
\texttt{torch.serialization.os.system}).
\modelscan fails to recognize this as malicious because the malicious callable
appears to be part of an allowed module, and the scanner does not inspect deeper;
\modeltracer successfully identifies this malicious behavior, but only after the
malicious call has been invoked and its underlying system call appears in the
execution trace (this is by design, as a dynamic scanning tool).

\section{\tool Implementation} \label{ImpAPdix}

In~\cref{SEC:IMPLEMENTATION} we summarized the implementation of \tool.
Here we provide more details.

\para{Integrated components}
We built \tool's policy generation using the Joern
framework~\cite{yamaguchi2014joern}, a static program
analysis tool which supports Python code. We use it to construct an \AST with recovered type information
for the analyzed library code. In our implementation, forked from Joern version
2.0.385, we found several limitations in Joern's type recovery pass for Python
features, which we reported and fixed, such as failures to:
  (1) track types of variables assigned to collection objects,
  (2) recover types of nested attributes,
  (3) recover type information without an assignment expression, even when provided type annotations,
  and
  (4) correctly record class inheritance information.
\tool's policy generation component is implemented as a Scala program that
interfaces with the Joern CPG and implements \cref{ALGO:STATIC_INFERENCE} by
querying for the required class properties.

We built \tool's policy enforcement component by modifying the existing
CPython implementation of the \picklemachine (commit a365dd6)~\cite{pickle-src-fork}
to take model loading policies as inputs
and to lazily load \pickle programs. We changed the semantics of the
\texttt{GLOBAL}, \texttt{STACK\_GLOBAL}, and \texttt{REDUCE} opcodes to respect
policy settings, removed the \texttt{OBJ} and \texttt{INST} legacy opcodes, and
removed support for extension codes, in line with best practices when securing
the \picklemachine~\cite{weightsOnlyUnpickler}. We also modified the
\texttt{BUILD} opcode to prevent altering the \texttt{\_\_name\_\_} and
\texttt{\_\_module\_\_} attributes of callables. We made no changes to the
interface to invoke our modified \picklemachine (\inlPython{pickle.load}) so
that it serves as a drop-in replacement for the unrestricted \picklemachine in
the CPython runtime.

\begin{figure*}[t]

    \centering
    \includegraphics[width=\textwidth]{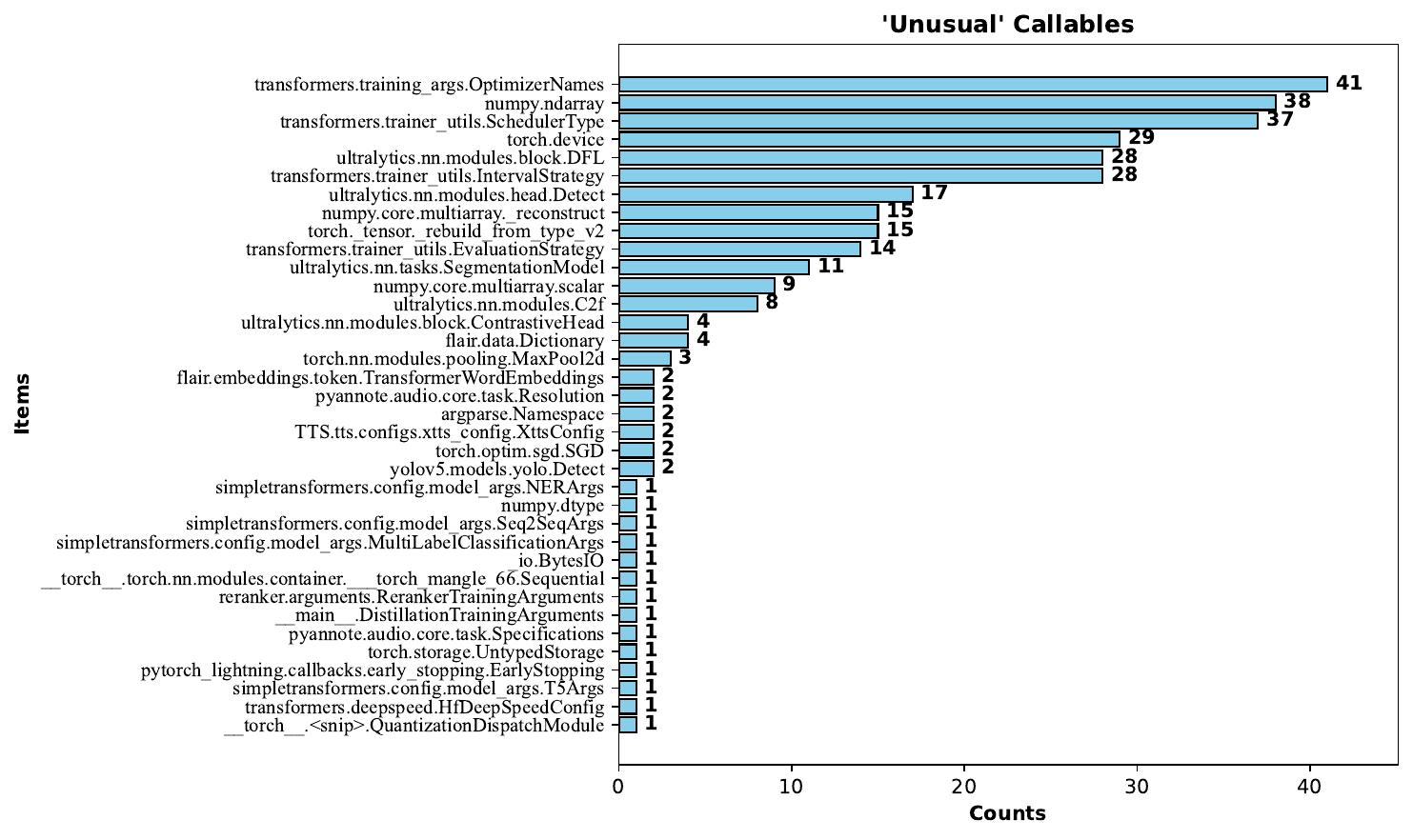}
        \caption{Callables observed in the \tracedRepositoriesDisallowed model
        repositories that cannot be loaded by the \weightsonly. A few callables
        from popular libraries (Hugging Face transformers, numpy) dominate, but
        a tail of callables come from small ML libraries (flair, fairseq,
        yolov5).
    }
    \label{FIG:UNUSUAL_CALLABLES}
\end{figure*}

\section{Libraries and Models in Benign Dataset}\label{SEC:APDX:DATASET}

\tool generates policies for ML libraries to correctly load models using those
libraries.
To evaluate this, we constructed a dataset of \ndatasetlibraries libraries and
\nsafedatasetmodels benign models sourced from Hugging Face (\S\ref{SEC:EVAL_DATASET}).
\cref{TABLE:library-versions} lists the libraries and identifying
information, including the git commit used during the evaluation.
\cref{TABLE:model-versions} lists the models and identifying information,
including the Hugging Face git commit for the model version that \tool loads.
Further information, like generated policies for each library and model load
success rates, are available in the accompanying software artifact.

\begin{NoDiff}

\onecolumn

\begin{table}[ht]
  \centering
  \caption{Source and commit information for all libraries in the \tool evaluation.
  A \checkmark~in the \textbf{Modified} column indicates that we performed manual
  modifications to the library source code before analysis, described in \cref{SEC:EVAL_DATASET}.}
  \label{TABLE:library-versions}
  \begin{tabular}{lclc}
    \toprule
    \textbf{Library} & \textbf{Modified} & \textbf{GitHub Repository} & \textbf{Git Commit} \\
    \cmidrule{1-4}
    \libraryconch & - & \libraryconchrepo & \libraryconchhash \\
    \libraryflagembedding & - & \libraryflagembeddingrepo & \libraryflagembeddinghash \\
    \libraryflair & \checkmark & \libraryflairrepo & \libraryflairhash \\
    \librarygliner & - & \libraryglinerrepo & \libraryglinerhash \\
    \libraryhuggingsound & - & \libraryhuggingsoundrepo & \libraryhuggingsoundhash \\
    \librarylanguagebind & - & \librarylanguagebindrepo & \librarylanguagebindhash \\
    \librarymelotts & - & \librarymelottsrepo & \librarymelottshash \\
    \libraryparrot & - & \libraryparrotrepo & \libraryparrothash \\
    \librarypyannote & \checkmark & \librarypyannoterepo & \librarypyannotehash \\
    \librarypysentimiento & - & \librarypysentimientorepo & \librarypysentimientohash \\
    \librarysentencetransformers & \checkmark & \librarysentencetransformersrepo & \librarysentencetransformershash \\
    \librarysuperimage & -  & \librarysuperimagerepo & \librarysuperimagehash \\
    \librarytner & - & \librarytnerrepo & \librarytnerhash \\
    \librarytweetnlp & - & \librarytweetnlprepo & \librarytweetnlphash \\
    \libraryyolovfive & \checkmark & \libraryyolovfiverepo & \libraryyolovfivehash \\
    \libraryyoloveleven & \checkmark & \libraryyolovelevenrepo & \libraryyolovelevenhash \\
    \bottomrule
  \end{tabular}
\end{table}

\begin{longtable}{@{}p{3.8cm}p{9cm}>{\centering\arraybackslash}p{3cm}@{}}
\caption{Source and version information for all benign models in the \tool
evaluation dataset (\S\ref{SEC:EVAL_DATASET}). When a single repository contains multiple \pickle models,
we list the repository once per model with the model name listed in parentheses.
The Git Commit refers to the Hugging Face repository commit containing the
version of the model in the dataset. Two models in the dataset were removed from
Hugging Face after the evaluation concluded and we
were unable to record their commit hashes; these are recorded as ``N/A''.} \label{TABLE:model-versions} \\
\toprule
\textbf{Library} & \textbf{Hugging Face Repository (Model Name)} & \textbf{Git Commit} \\
\midrule
\endfirsthead

\toprule
\textbf{Library} & \textbf{Hugging Face Repository (Model Name)} & \textbf{Git Commit} \\
\midrule
\endhead

\midrule
\endfoot

\bottomrule
\endlastfoot

\libraryconch & MahmoodLab/CONCH & \texttt{f9ca9f8} \\
\libraryflagembedding & BAAI/bge-base-en & \texttt{b737bf5} \\
\libraryflagembedding & BAAI/bge-base-en-v1.5 & \texttt{a5beb1e} \\
\libraryflagembedding & BAAI/bge-base-zh-v1.5 & \texttt{f03589c} \\
\libraryflagembedding & BAAI/bge-large-en & \texttt{abe7d9d} \\
\libraryflagembedding & BAAI/bge-large-en-v1.5 & \texttt{d4aa690} \\
\libraryflagembedding & BAAI/bge-large-zh & \texttt{b5d9f5c} \\
\libraryflagembedding & BAAI/bge-large-zh-v1.5 & \texttt{79e7739} \\
\libraryflagembedding & BAAI/bge-reranker-base & \texttt{2cfc18c} \\
\libraryflagembedding & BAAI/bge-reranker-large & \texttt{55611d7} \\
\libraryflagembedding & BAAI/bge-small-en & \texttt{2275a7b} \\
\libraryflagembedding & BAAI/bge-small-en-v1.5 & \texttt{5c38ec7} \\
\libraryflagembedding & BAAI/bge-small-zh & \texttt{1d2363c} \\
\libraryflagembedding & BAAI/bge-small-zh-v1.5 & \texttt{7999e1d} \\
\libraryflagembedding & BAAI/llm-embedder & \texttt{c3e8ac8} \\
\libraryflair & flair/chunk-english-fast & \texttt{34cdda2} \\
\libraryflair & flair/ner-dutch-large & \texttt{44c2859} \\
\libraryflair & flair/ner-english & \texttt{b13c26b} \\
\libraryflair & flair/ner-english-fast & \texttt{f75577b} \\
\libraryflair & flair/ner-english-large & \texttt{e2b1caa} \\
\libraryflair & flair/ner-english-ontonotes & \texttt{ffa7600} \\
\libraryflair & flair/ner-english-ontonotes-fast & \texttt{4f31790} \\
\libraryflair & flair/ner-english-ontonotes-large & \texttt{4ffb359} \\
\libraryflair & flair/ner-french & \texttt{27fb2ba} \\
\libraryflair & flair/ner-german & \texttt{4e3f3d1} \\
\libraryflair & flair/ner-german-large & \texttt{4b459fa} \\
\libraryflair & flair/ner-multi & \texttt{9e5dc17} \\
\libraryflair & flair/ner-spanish-large & \texttt{9d4671d} \\
\libraryflair & flair/pos-english & \texttt{b32242e} \\
\libraryflair & flair/pos-english-fast & \texttt{78bf413} \\
\libraryflair & flair/upos-english & \texttt{dbd8c36} \\
\libraryflair & flair/upos-english-fast & \texttt{8748c22} \\
\libraryflair & flair/upos-multi & \texttt{0ee3d86} \\
\librarygliner & DeepMount00/GLiNER\_ITA\_BASE & \texttt{8dadd43} \\
\librarygliner & DeepMount00/GLiNER\_ITA\_LARGE & \texttt{7f74be0} \\
\librarygliner & DeepMount00/GLiNER\_PII\_ITA & \texttt{b31700c} \\
\librarygliner & DeepMount00/universal\_ner\_ita & \texttt{158e11d} \\
\librarygliner & EmergentMethods/gliner\_medium\_news-v2.1 & \texttt{2450430} \\
\librarygliner & gliner-community/gliner\_large-v2.5 & \texttt{3b3bcae} \\
\librarygliner & gliner-community/gliner\_medium-v2.5 & \texttt{ed16f26} \\
\librarygliner & numind/NuNER\_Zero & \texttt{9c23c20} \\
\librarygliner & urchade/gliner\_base & \texttt{b448aaf} \\
\librarygliner & urchade/gliner\_large-v1 & \texttt{1f55b52} \\
\librarygliner & urchade/gliner\_large-v2 & \texttt{416a9b8} \\
\librarygliner & urchade/gliner\_large-v2.1 & \texttt{abd49a1} \\
\librarygliner & urchade/gliner\_medium-v2.1 & \texttt{a3f776a} \\
\librarygliner & urchade/gliner\_multi & \texttt{b5720ab} \\
\librarygliner & urchade/gliner\_multi-v2.1 & \texttt{853ce23} \\
\librarygliner & urchade/gliner\_multi\_pii-v1 & \texttt{1fcf13e} \\
\librarygliner & urchade/gliner\_small-v2.1 & \texttt{4e09141} \\
\libraryhuggingsound & jonatasgrosman/wav2vec2-large-xlsr-53-arabic & \texttt{af46c2d} \\
\libraryhuggingsound & jonatasgrosman/wav2vec2-large-xlsr-53-chinese-zh-cn & \texttt{99ccb27} \\
\libraryhuggingsound & jonatasgrosman/wav2vec2-large-xlsr-53-dutch & \texttt{46f2213} \\
\libraryhuggingsound & jonatasgrosman/wav2vec2-large-xlsr-53-english & \texttt{569a623} \\
\libraryhuggingsound & jonatasgrosman/wav2vec2-large-xlsr-53-finnish & \texttt{a497f86} \\
\libraryhuggingsound & jonatasgrosman/wav2vec2-large-xlsr-53-french & \texttt{7c79e10} \\
\libraryhuggingsound & jonatasgrosman/wav2vec2-large-xlsr-53-german & \texttt{4b8a029} \\
\libraryhuggingsound & jonatasgrosman/wav2vec2-large-xlsr-53-greek & \texttt{489b34f} \\
\libraryhuggingsound & jonatasgrosman/wav2vec2-large-xlsr-53-hungarian & \texttt{2bd0786} \\
\libraryhuggingsound & jonatasgrosman/wav2vec2-large-xlsr-53-italian & \texttt{dab04a3} \\
\libraryhuggingsound & jonatasgrosman/wav2vec2-large-xlsr-53-japanese & \texttt{cf031e0} \\
\libraryhuggingsound & jonatasgrosman/wav2vec2-large-xlsr-53-persian & \texttt{2347140} \\
\libraryhuggingsound & jonatasgrosman/wav2vec2-large-xlsr-53-polish & \texttt{6b1cea3} \\
\libraryhuggingsound & jonatasgrosman/wav2vec2-large-xlsr-53-portuguese & \texttt{634ac65} \\
\libraryhuggingsound & jonatasgrosman/wav2vec2-large-xlsr-53-russian & \texttt{2329100} \\
\libraryhuggingsound & jonatasgrosman/wav2vec2-large-xlsr-53-spanish & \texttt{96d7e9b} \\
\libraryhuggingsound & jonatasgrosman/wav2vec2-xls-r-1b-portuguese & \texttt{8926743} \\
\librarylanguagebind & LanguageBind/LanguageBind\_Audio & \texttt{7aea390} \\
\librarylanguagebind & LanguageBind/LanguageBind\_Audio\_FT & \texttt{4820c49} \\
\librarylanguagebind & LanguageBind/LanguageBind\_Image & \texttt{d8c2e37} \\
\librarylanguagebind & LanguageBind/LanguageBind\_Video & \texttt{84bed6c} \\
\librarylanguagebind & LanguageBind/LanguageBind\_Video\_FT & \texttt{13f52c2} \\
\librarylanguagebind & LanguageBind/LanguageBind\_Video\_Huge\_V1.5\_FT & \texttt{dd4bbe0} \\
\librarylanguagebind & LanguageBind/LanguageBind\_Video\_merge & \texttt{efc40ec} \\
\librarylanguagebind & LanguageBind/LanguageBind\_Video\_V1.5\_FT & \texttt{5d53aab} \\
\librarymelotts & myshell-ai/MeloTTS-Chinese & \texttt{af5d207} \\
\librarymelotts & myshell-ai/MeloTTS-English & \texttt{bb4fb73} \\
\librarymelotts & myshell-ai/MeloTTS-English-v2 & \texttt{a53e350} \\
\librarymelotts & myshell-ai/MeloTTS-English-v3 & \texttt{f7c4a35} \\
\librarymelotts & myshell-ai/MeloTTS-French & \texttt{1e9bf59} \\
\librarymelotts & myshell-ai/MeloTTS-Japanese & \texttt{367f879} \\
\librarymelotts & myshell-ai/MeloTTS-Korean & \texttt{0207e5a} \\
\librarymelotts & myshell-ai/MeloTTS-Spanish & \texttt{dbb5496} \\
\libraryparrot & prithivida/parrot\_paraphraser\_on\_T5 & \texttt{9f32aa1} \\
\librarypyannote & collinbarnwell/pyannote-segmentation-30 & \texttt{f47575f} \\
\librarypyannote & EonNextPlatform/pyannote-wespeaker-voxceleb-resnet34-LM & N/A \\
\librarypyannote & EonNextPlatform/segmentation-3.0 & N/A \\
\librarypyannote & fatymatariq/pyannote-wespeaker-voxceleb-resnet34-LM & \texttt{27e7027} \\
\librarypyannote & fatymatariq/segmentation-3.0 & \texttt{e1b9697} \\
\librarypyannote & philschmid/pyannote-segmentation & \texttt{d13283c} \\
\librarypyannote & pyannote/brouhaha & \texttt{c93c9b5} \\
\librarypyannote & pyannote/embedding & \texttt{4db4899} \\
\librarypyannote & pyannote/segmentation & \texttt{660b9e2} \\
\librarypyannote & pyannote/segmentation-3.0 & \texttt{e66f3d3} \\
\librarypyannote & pyannote/wespeaker-voxceleb-resnet34-LM & \texttt{837717d} \\
\librarypyannote & Revai/reverb-diarization-v1 & \texttt{4ad4567} \\
\librarypyannote & Revai/reverb-diarization-v2 & \texttt{f086648} \\
\librarypyannote & tensorlake/segmentation-3.0 & \texttt{035d994} \\
\librarypysentimiento & pysentimiento/bertweet-pt-sentiment & \texttt{7266128} \\
\librarypysentimiento & pysentimiento/robertuito-ner & \texttt{43dde63} \\
\librarypysentimiento & pysentimiento/robertuito-sentiment-analysis & \texttt{a2cc0f6} \\
\librarysentencetransformers & avsolatorio/GIST-all-MiniLM-L6-v2 & \texttt{ea89dfa} \\
\librarysentencetransformers & avsolatorio/GIST-Embedding-v0 & \texttt{bf6b2e5} \\
\librarysentencetransformers & avsolatorio/GIST-small-Embedding-v0 & \texttt{d6c4190} \\
\librarysentencetransformers & BAAI/bge-base-en & \texttt{b737bf5} \\
\librarysentencetransformers & BAAI/bge-base-en-v1.5 & \texttt{a5beb1e} \\
\librarysentencetransformers & BAAI/bge-large-en & \texttt{abe7d9d} \\
\librarysentencetransformers & BAAI/bge-large-en-v1.5 & \texttt{d4aa690} \\
\librarysentencetransformers & BAAI/bge-large-zh-v1.5 & \texttt{79e7739} \\
\librarysentencetransformers & BAAI/bge-reranker-base & \texttt{2cfc18c} \\
\librarysentencetransformers & BAAI/bge-reranker-large & \texttt{55611d7} \\
\librarysentencetransformers & BAAI/bge-small-en & \texttt{2275a7b} \\
\librarysentencetransformers & BAAI/bge-small-en-v1.5 & \texttt{5c38ec7} \\
\librarysentencetransformers & BAAI/llm-embedder & \texttt{c3e8ac8} \\
\librarysentencetransformers & cointegrated/rubert-tiny2 & \texttt{dad72b8} \\
\librarysentencetransformers & DMetaSoul/sbert-chinese-general-v2 & \texttt{14b486c} \\
\librarysentencetransformers & flax-sentence-embeddings/all\_datasets\_v4\_MiniLM-L6 & \texttt{a407cc0} \\
\librarysentencetransformers & hiiamsid/sentence\_similarity\_spanish\_es & \texttt{3118431} \\
\librarysentencetransformers & intfloat/e5-base-v2 & \texttt{1c644c9} \\
\librarysentencetransformers & intfloat/e5-large-v2 & \texttt{b322e09} \\
\librarysentencetransformers & intfloat/e5-mistral-7b-instruct & \texttt{07163b7} \\
\librarysentencetransformers & intfloat/e5-small-v2 & \texttt{dca8b1a} \\
\librarysentencetransformers & intfloat/multilingual-e5-base & \texttt{d13f1b2} \\
\librarysentencetransformers & intfloat/multilingual-e5-large & \texttt{ab10c1a} \\
\librarysentencetransformers & intfloat/multilingual-e5-small & \texttt{fd1525a} \\
\librarysentencetransformers & jhgan/ko-sbert-nli & \texttt{b78c95e} \\
\librarysentencetransformers & jhgan/ko-sroberta-multitask & \texttt{ab957ae} \\
\librarysentencetransformers & naufalihsan/indonesian-sbert-large & \texttt{a5cbfbd} \\
\librarysentencetransformers & NeuML/pubmedbert-base-embeddings & \texttt{ba210f4} \\
\librarysentencetransformers & nomic-ai/nomic-embed-text-v1 & \texttt{cc62377} \\
\librarysentencetransformers & pritamdeka/BioBERT-mnli-snli-scinli-scitail-mednli-stsb & \texttt{82d4468} \\
\librarysentencetransformers & pritamdeka/S-PubMedBert-MS-MARCO & \texttt{96786c7} \\
\librarysentencetransformers & sentence-transformers-testing/stsb-bert-tiny-safetensors & \texttt{f3cb857} \\
\librarysentencetransformers & sentence-transformers/all-distilroberta-v1 & \texttt{8d88b92} \\
\librarysentencetransformers & sentence-transformers/all-MiniLM-L12-v2 & \texttt{364dd28} \\
\librarysentencetransformers & sentence-transformers/all-MiniLM-L6-v2 & \texttt{fa97f6e} \\
\librarysentencetransformers & sentence-transformers/all-mpnet-base-v2 & \texttt{9a32259} \\
\librarysentencetransformers & sentence-transformers/all-roberta-large-v1 & \texttt{c8b9f2a} \\
\librarysentencetransformers & sentence-transformers/bert-base-nli-mean-tokens & \texttt{2511498} \\
\librarysentencetransformers & sentence-transformers/bert-base-nli-stsb-mean-tokens & \texttt{767076a} \\
\librarysentencetransformers & sentence-transformers/clip-ViT-B-32-multilingual-v1 & \texttt{58edf8c} \\
\librarysentencetransformers & sentence-transformers/distilbert-base-nli-mean-tokens & \texttt{5aa678b} \\
\librarysentencetransformers & sentence-transformers/distilbert-base-nli-stsb-mean-tokens & \texttt{cb8a28f} \\
\librarysentencetransformers & sentence-transformers/distilbert-multilingual-nli-stsb-quora-ranking & \texttt{57cf088} \\
\librarysentencetransformers & sentence-transformers/distiluse-base-multilingual-cased-v1 & \texttt{457e815} \\
\librarysentencetransformers & sentence-transformers/distiluse-base-multilingual-cased-v2 & \texttt{dad0fa1} \\
\librarysentencetransformers & sentence-transformers/LaBSE & \texttt{b7f9471} \\
\librarysentencetransformers & sentence-transformers/msmarco-bert-base-dot-v5 & \texttt{c45bf94} \\
\librarysentencetransformers & sentence-transformers/msmarco-distilbert-base-dot-prod-v3 & \texttt{0cf6cf1} \\
\librarysentencetransformers & sentence-transformers/msmarco-distilbert-base-tas-b & \texttt{136d171} \\
\librarysentencetransformers & sentence-transformers/msmarco-distilbert-base-v4 & \texttt{19f0f4c} \\
\librarysentencetransformers & sentence-transformers/msmarco-distilbert-cos-v5 & \texttt{c598d92} \\
\librarysentencetransformers & sentence-transformers/msmarco-distilbert-dot-v5 & \texttt{6ad1718} \\
\librarysentencetransformers & sentence-transformers/msmarco-MiniLM-L-6-v3 & \texttt{d273900} \\
\librarysentencetransformers & sentence-transformers/msmarco-MiniLM-L12-cos-v5 & \texttt{09660d8} \\
\librarysentencetransformers & sentence-transformers/msmarco-MiniLM-L6-cos-v5 & \texttt{14ca9be} \\
\librarysentencetransformers & sentence-transformers/multi-qa-distilbert-cos-v1 & \texttt{bc2339d} \\
\librarysentencetransformers & sentence-transformers/multi-qa-MiniLM-L6-cos-v1 & \texttt{b207367} \\
\librarysentencetransformers & sentence-transformers/multi-qa-mpnet-base-cos-v1 & \texttt{822dbc9} \\
\librarysentencetransformers & sentence-transformers/multi-qa-mpnet-base-dot-v1 & \texttt{4633e80} \\
\librarysentencetransformers & sentence-transformers/nli-mpnet-base-v2 & \texttt{688eb0a} \\
\librarysentencetransformers & sentence-transformers/paraphrase-albert-small-v2 & \texttt{39d5b65} \\
\librarysentencetransformers & sentence-transformers/paraphrase-MiniLM-L12-v2 & \texttt{3f21b01} \\
\librarysentencetransformers & sentence-transformers/paraphrase-MiniLM-L3-v2 & \texttt{029a79d} \\
\librarysentencetransformers & sentence-transformers/paraphrase-MiniLM-L6-v2 & \texttt{9a27583} \\
\librarysentencetransformers & sentence-transformers/paraphrase-mpnet-base-v2 & \texttt{bef3689} \\
\librarysentencetransformers & sentence-transformers/paraphrase-multilingual-MiniLM-L12-v2 & \texttt{8d6b950} \\
\librarysentencetransformers & sentence-transformers/paraphrase-multilingual-mpnet-base-v2 & \texttt{75c5775} \\
\librarysentencetransformers & sentence-transformers/sentence-t5-base & \texttt{50c53e2} \\
\librarysentencetransformers & sentence-transformers/sentence-t5-xl & \texttt{2965d31} \\
\librarysentencetransformers & sentence-transformers/stsb-roberta-base & \texttt{fb8c0e7} \\
\librarysentencetransformers & sentence-transformers/stsb-xlm-r-multilingual & \texttt{18f85ee} \\
\librarysentencetransformers & shibing624/text2vec-base-chinese & \texttt{183bb99} \\
\librarysentencetransformers & shibing624/text2vec-base-multilingual & \texttt{6633dc4} \\
\librarysentencetransformers & Supabase/gte-small & \texttt{93b36ff} \\
\librarysentencetransformers & thenlper/gte-base & \texttt{c078288} \\
\librarysentencetransformers & thenlper/gte-large & \texttt{4bef63f} \\
\librarysentencetransformers & thenlper/gte-small & \texttt{17e1f34} \\
\librarysuperimage & eugenesiow/edsr (pytorch\_model\_2x.pt) & \texttt{8f214a5} \\
\librarysuperimage & eugenesiow/edsr (pytorch\_model\_3x.pt) & \texttt{8f214a5} \\
\librarysuperimage & eugenesiow/edsr (pytorch\_model\_4x.pt) & \texttt{8f214a5} \\
\librarysuperimage & eugenesiow/edsr-base (pytorch\_model\_2x.pt) & \texttt{d622f68} \\
\librarysuperimage & eugenesiow/edsr-base (pytorch\_model\_3x.pt) & \texttt{d622f68} \\
\librarysuperimage & eugenesiow/edsr-base (pytorch\_model\_4x.pt) & \texttt{d622f68} \\
\librarytner & tner/deberta-v3-large-ontonotes5 & \texttt{98a5818} \\
\librarytner & tner/roberta-large-ontonotes5 & \texttt{0bce50f} \\
\librarytner & tner/roberta-large-tweetner7-all & \texttt{e7fbeec} \\
\librarytner & tner/roberta-large-wnut2017 & \texttt{e2e15e4} \\
\librarytweetnlp & cardiffnlp/twitter-xlm-roberta-base-sentiment-multilingual & \texttt{82107f4} \\
\libraryyolovfive & keremberke/yolov5m-smoke & \texttt{79dc392} \\
\libraryyolovfive & keremberke/yolov5m-license-plate & \texttt{2bc5d84} \\
\libraryyolovfive & keremberke/yolov5n-license-plate & \texttt{b9a03ec} \\
\libraryyolovfive & Ultralytics/YOLOv5 (yolov5s.pt) & \texttt{73fb27f} \\
\libraryyolovfive & Ultralytics/YOLOv5 (yolov5s6.pt) & \texttt{73fb27f} \\
\libraryyolovfive & Ultralytics/YOLOv5 (yolov5n.pt) & \texttt{73fb27f} \\
\libraryyolovfive & Ultralytics/YOLOv5 (yolov5n6.pt) & \texttt{73fb27f} \\
\libraryyolovfive & Ultralytics/YOLOv5 (yolov5m.pt) & \texttt{73fb27f} \\
\libraryyolovfive & Ultralytics/YOLOv5 (yolov5m6.pt) & \texttt{73fb27f} \\
\libraryyolovfive & Ultralytics/YOLOv5 (yolov5l.pt) & \texttt{73fb27f} \\
\libraryyolovfive & Ultralytics/YOLOv5 (yolov5l6.pt) & \texttt{73fb27f} \\
\libraryyolovfive & keremberke/yolov5s-license-plate & \texttt{038440e} \\
\libraryyoloveleven & Bingsu/yolo-world-mirror (yolov8l-world.pt) & \texttt{414d0ee} \\
\libraryyoloveleven & Bingsu/yolo-world-mirror (yolov8l-worldv2.pt) & \texttt{414d0ee} \\
\libraryyoloveleven & Bingsu/yolo-world-mirror (yolov8m-world.pt) & \texttt{414d0ee} \\
\libraryyoloveleven & Bingsu/yolo-world-mirror (yolov8m-worldv2.pt) & \texttt{414d0ee} \\
\libraryyoloveleven & Bingsu/yolo-world-mirror (yolov8s-world.pt) & \texttt{414d0ee} \\
\libraryyoloveleven & Bingsu/yolo-world-mirror (yolov8s-worldv2.pt) & \texttt{414d0ee} \\
\libraryyoloveleven & Bingsu/yolo-world-mirror (yolov8x-world.pt) & \texttt{414d0ee} \\
\libraryyoloveleven & Bingsu/yolo-world-mirror (yolov8x-worldv2.pt) & \texttt{414d0ee} \\
\libraryyoloveleven & keremberke/yolov8m-table-extraction & \texttt{8826513} \\
\libraryyoloveleven & keremberke/yolov8s-table-extraction & \texttt{fc8bf12} \\
\libraryyoloveleven & Ultralytics/YOLOv8 (yolov8l.pt) & \texttt{5d9ba66} \\
\libraryyoloveleven & Ultralytics/YOLOv8 (yolov8m.pt) & \texttt{5d9ba66} \\
\libraryyoloveleven & Ultralytics/YOLOv8 (yolov8n.pt) & \texttt{5d9ba66} \\
\libraryyoloveleven & Ultralytics/YOLOv8 (yolov8s.pt) & \texttt{5d9ba66} \\
\libraryyoloveleven & Ultralytics/YOLOv8 (yolov8x.pt) & \texttt{5d9ba66} \\
\libraryyoloveleven & Ultralytics/YOLO11 (yolo11l-pose.pt) & \texttt{9cc3192} \\
\libraryyoloveleven & Ultralytics/YOLO11 (yolo11l.pt) & \texttt{9cc3192} \\
\libraryyoloveleven & Ultralytics/YOLO11 (yolo11l-seg.pt) & \texttt{9cc3192} \\
\libraryyoloveleven & Ultralytics/YOLO11 (yolo11m-pose.pt) & \texttt{9cc3192} \\
\libraryyoloveleven & Ultralytics/YOLO11 (yolo11m.pt) & \texttt{9cc3192} \\
\libraryyoloveleven & Ultralytics/YOLO11 (yolo11m-seg.pt) & \texttt{9cc3192} \\
\libraryyoloveleven & Ultralytics/YOLO11 (yolo11n-pose.pt) & \texttt{9cc3192} \\
\libraryyoloveleven & Ultralytics/YOLO11 (yolo11n.pt) & \texttt{9cc3192} \\
\libraryyoloveleven & Ultralytics/YOLO11 (yolo11n-seg.pt) & \texttt{9cc3192} \\
\libraryyoloveleven & Ultralytics/YOLO11 (yolo11s-pose.pt) & \texttt{9cc3192} \\
\libraryyoloveleven & Ultralytics/YOLO11 (yolo11s.pt) & \texttt{9cc3192} \\
\libraryyoloveleven & Ultralytics/YOLO11 (yolo11s-seg.pt) & \texttt{9cc3192} \\
\libraryyoloveleven & Ultralytics/YOLO11 (yolo11x-pose.pt) & \texttt{9cc3192} \\
\libraryyoloveleven & Ultralytics/YOLO11 (yolo11x.pt) & \texttt{9cc3192} \\
\libraryyoloveleven & Ultralytics/YOLO11 (yolo11x-seg.pt) & \texttt{9cc3192} \\
\libraryyoloveleven & Anzhc/Anzhcs\_YOLOs (Anzhc Breasts Seg v1 1024m.pt) & \texttt{4bb7d18} \\
\libraryyoloveleven & Anzhc/Anzhcs\_YOLOs (Anzhc Breasts Seg v1 1024n.pt) & \texttt{4bb7d18} \\
\libraryyoloveleven & Anzhc/Anzhcs\_YOLOs (Anzhc Breasts Seg v1 1024s.pt) & \texttt{4bb7d18} \\
\libraryyoloveleven & Anzhc/Anzhcs\_YOLOs (Anzhc Eyes -seg-hd.pt) & \texttt{4bb7d18} \\
\libraryyoloveleven & Anzhc/Anzhcs\_YOLOs (Anzhc Face seg 1024 v2 y8n.pt) & \texttt{4bb7d18} \\
\libraryyoloveleven & Anzhc/Anzhcs\_YOLOs (Anzhc Face seg 640 v2 y8n.pt) & \texttt{4bb7d18} \\
\libraryyoloveleven & Anzhc/Anzhcs\_YOLOs (Anzhc Face seg 640 v3 y11n.pt) & \texttt{4bb7d18} \\
\libraryyoloveleven & Anzhc/Anzhcs\_YOLOs (Anzhc Face seg 768MS v2 y8n.pt) & \texttt{4bb7d18} \\
\libraryyoloveleven & Anzhc/Anzhcs\_YOLOs (Anzhc Face seg 768 v2 y8n.pt) & \texttt{4bb7d18} \\
\libraryyoloveleven & Anzhc/Anzhcs\_YOLOs (Anzhc Face -seg.pt) & \texttt{4bb7d18} \\
\libraryyoloveleven & Anzhc/Anzhcs\_YOLOs (Anzhc HeadHair seg y8m.pt) & \texttt{4bb7d18} \\
\libraryyoloveleven & Anzhc/Anzhcs\_YOLOs (Anzhc HeadHair seg y8n.pt) & \texttt{4bb7d18} \\
\libraryyoloveleven & Anzhc/Anzhcs\_YOLOs (Anzhc Manga Panels -seg.pt) & \texttt{4bb7d18} \\
\libraryyoloveleven & Anzhc/Anzhcs\_YOLOs (Anzhcs Drones v03 1024 y11n.pt) & \texttt{4bb7d18} \\
\libraryyoloveleven & Anzhc/Anzhcs\_YOLOs (Anzhcs ManFace v02 1024 y8n.pt) & \texttt{4bb7d18} \\
\libraryyoloveleven & Anzhc/Anzhcs\_YOLOs (Anzhcs WomanFace v05 1024 y8n.pt) & \texttt{4bb7d18} \\
\libraryyoloveleven & tech4humans/yolov8s-signature-detector & \texttt{c891f02} \\
\libraryyoloveleven & foduucom/stockmarket-pattern-detection-yolov8 & \texttt{31bbb9a} \\
\libraryyoloveleven & pitangent-ds/YOLOv8-human-detection-thermal & \texttt{a5d30f1} \\
\libraryyoloveleven & arnabdhar/YOLOv8-nano-aadhar-card & \texttt{d5f938d} \\
\libraryyoloveleven & keremberke/yolov8n-building-segmentation & \texttt{eb61328} \\

\end{longtable}

\twocolumn %
\end{NoDiff}

\else
\fi

\end{document}